\documentclass[aps,twocolumn,secnumarabic,nobalancelastpage,amsmath,amssymb,nofootinbib,preprintnumbers]{revtex4-1}

\pdfoutput=1

\usepackage{slashed}
\usepackage{graphics}
\usepackage{graphicx}
\usepackage{caption}
\usepackage{subcaption}
\usepackage{url}

\begin{document}

\preprint{UCI-HEP-TR-2013-17}
\preprint{KEK-TH 1659}

\title{Simplified Models for Dark Matter Interacting with Quarks}
\author{Anthony DiFranzo$^1$}
\author{Keiko I. Nagao$^{2,3}$}
\author{Arvind Rajaraman$^1$}
\author{Tim M.P. Tait$^1$}
\affiliation{$^1$Department of Physics and Astronomy, University of California, Irvine, CA 92697}
\affiliation{$^2$KEK Theory Center, Tsukuba, Ibaraki 305-0801, Japan}
\affiliation{$^3$Department of Engineering Science, Niihama National College of Technology, Niihama, Ehime 792-8580, Japan}
\date{\today}

\begin{abstract}
We investigate simplified models in which dark matter particles, taken to be either Dirac or Majorana fermions,
couple to quarks via colored mediators.  We determine bounds from colliders and direct detection experiments,
and show how the interplay of the two leads to a complementary view of this class of dark matter models.
Forecasts for future searches in light of the current constraints are presented.
\end{abstract}

\maketitle

\section{Introduction}

The largest mass component of the universe is a longstanding mystery to the physics community. Dark Matter has been probed using particle colliders, direct detection experiments, and indirect detection in telescopes. Despite these intensive searches it has proven elusive. It is clear we must be thorough and creative as we continue the important mission to search for it.

From a particle physics perspective, it is essential to have a theoretical description of dark matter.  Such a description allows one to contrast various searches \cite{Bauer:2013ihz}, and fit the dark matter into the broader context of the Standard Model (SM).
There are two common approaches to dark matter model building. The first is to find a complete theory (which typically addresses some other problem such as the gauge hierarchy),
and also contains a potential dark matter candidate. The prototypical example of this approach is the
Minimal Supersymmetric Standard Model (MSSM). One can work with the complete theory with every detail included. Often there are many free parameters, which on the one hand leads to a rich range of phenomena, but on the other makes it difficult to draw general conclusions (see, e.g. \cite{Cahill-Rowley:2013dpa}). An alternative is to look at Simplified models \cite{Alves:2011wf} where only the particles and parameters most relevant to the search are included. The simplified model can be understood as a phenomenological sketch of a complete model; for example, in the MSSM this might include only looking at the two lightest supersymmetric particles.

The most model-independent theory of dark matter is to adopt an Effective Field Theory (EFT) which includes
only the SM plus the dark matter itself as states in the theory.  This approach has led to a fruitful program,
in which one enumerates the possible operators describing dark matter interaction and places bounds on each of them (or in combination) from the null results of collider \cite{Beltran:2010ww} and/or astro-particle \cite{Beltran:2008xg} searches. This method's greatest power is that it is very generic. However it suffers from its UV incompleteness and breaks down at energies comparable to the mass of the underlying mediator particle, which can call its applicability into question, particularly at high energy colliders.

The strategy which we employ in this paper attempts to mediate these issues. A UV complete theory can be formed which is not necessarily a simplification of a particular complete theory, but instead captures features of classes of models.  The Simplified models we construct are broad enough to capture features of a certain class of models, while remaining valid at higher energies. Thus, it facilitates relating processes at varying energies, in particular results from collider and direct detection experiments. Specifically, we will examine models involving fermionic dark matter and scalar mediators.  For other constructions in similar directions, see 
Refs.~\cite{Goodman:2011jq}.

\section{Simplified Models}

Our simplified models contain a fermionic dark matter particle denoted by $\chi$, which 
is a SM singlet and can be either a Dirac or a Majorana fermion.  In addition, our theory contains scalar
mediator particles which interact with the dark matter and SM quarks.  Given our choice of dark matter
which is a SM singlet, this leads to three choices for gauge representations of the mediators
under the SM $( SU(3), SU(2) )_Y$:
\begin{eqnarray}
(3,1)_{2/3}, ~~~~~ (3,1)_{-1/3}, ~~~~~ (3,2)_{-1/6}.
\end{eqnarray}
We refer to these three choices as the $u_R$ model (with mediators labeled as $\tilde{u}$),
the $d_R$ model (with mediators $\tilde{d}$), and the $q_L$ model
(with mediators $\tilde{q}$).  For each model, the Lagangrian density consists of kinetic terms
for $\chi$ and the scalar mediators (including their gauge interactions), and a trilinear interaction
between $\chi$, the scalar mediator, and the corresponding SM quark.  For example, in the
$u_R$ model with Dirac DM the Lagrangian density is:
\begin{equation}\begin{split}
\mathcal{L} = i\bar{\chi}\slashed{\partial}\chi - M_{\chi}\bar{\chi}\chi + (D_{\mu}\tilde{u})^* (D^{\mu}\tilde{u}) - M_{\tilde{u}}^2\tilde{u}^* \tilde{u}\\ + (g_{\!_{DM}}\tilde{u}^* \bar{\chi}P_R u + h.c.)
\end{split}\end{equation}
where,
\begin{gather}
D_{\mu} \equiv \partial_{\mu} - ig_sG^a_{\mu}T^a - i\frac{2}{3}eA_{\mu},\\
u \equiv \begin{pmatrix} u\\c\\t \end{pmatrix},~~~~~
\tilde{u} \equiv \begin{pmatrix} \tilde{u}\\\tilde{c}\\\tilde{t} \end{pmatrix},
\end{gather}
such that $D_{\mu}$
defines the appropriate covariant derivative for the scalars, $u$ is the flavor
vector of up-type quarks, and $\tilde{u}$ a flavor vector of scalars.  Similar actions may be simply written
down for the $d_R$ and $q_L$ models, and the kinetic terms for $\chi$ may be appropriately modified
for the Majorana cases.

Each of our models is constructed with three scalar mediators, in order to implement 
Minimal Flavor Violation (MFV)  \cite{D'Ambrosio:2002ex}, with the mediators transforming
as triplets under the appropriate factors of the flavor group. This insures that in moving from the
gauge to mass basis, the deviations from flavor-universality will be proportional to SM Yukawa interactions, and
thus small for the first two generations.  In particular, this implies that the leading term (in an expansion in the
SM Yukawa matrices $Y^u$ and $Y^d$) $g_{DM}$ will be equal for all flavors, and that the masses 
$M_{\tilde{u}}^2$ will be degenerate.  We can further rephase $\chi$ such that $g_{DM}$ is a real parameter.

While we will not move beyond these leading terms in our analysis,
one can easily determine the next terms one would find (continuing to illustrate with the $u_R$ model):
\begin{eqnarray}
\mathcal{L}_{FV} & = &  
\left(
\delta g_{\!_{DM}} ~ \tilde{u}^* Y^{u\dag}Y^u \bar{\chi}P_R u + h.c.
\right)
 \nonumber \\ & & 
 ~~~~~~~ + \delta m^2 ~\tilde{u}^* Y^{u\dag}Y^u \tilde{u} + \mathcal{O}(Y^4)~.
 \label{eq:LFV}
\end{eqnarray}
The effect of non-zero $\delta g_{\!_{DM}}$ and/or $\delta m^2$ would be to split apart the couplings and
masses of the third generation mediators apart from the still largely degenerate first and second generations.
We defer consideration of non-zero $\delta g_{\!_{DM}}$ and $\delta m^2$ to future work.

\section{Experimental Constraints}

\subsection{Collider Bounds}

We derive bounds from CMS results which studied the simplfied model T2 \cite{CMS:zxa}, which assumes
that the dominant production mode produces two squark-like objects each of which subsequently decays into a jet and a neutralino. They place an upper limit on the production cross section of that process as a function of the masses of the squark and neutralino. For our corresponding process, two scalars are produced and, similarly, each decay into a jet and dark matter particle. The CMS study was conducted at $\sqrt{s} = 8$ TeV with 11.7 fb$^{-1}$ of data. The selection for this study requires 2-4 jets with $p_T>50$ GeV and $|\eta|<3.0$, where the two most energetic jets have $p_T>100$ GeV and $|\eta|<2.5$. Events with an isolated electron or muon with $p_T>10$ GeV and events with an isolated photon with $p_T>25$ GeV are vetoed. Events are also required to have $H_T>275$ GeV and $\alpha_T>0.55$. See \cite{CMS:zxa} for more detailed information.

We simulate production cross sections (including the decays) with used MadGraph 1.5.9 \cite{Alwall:2011uj} with model files implemented in FeynRules 1.7 \cite{Christensen:2008py}. We simulate at $\sqrt{s} = 8$ TeV
and using the CTEQ6L pdf set.  We have verified that we can reproduce the CMS limits for QCD production of
light squarks, up to the fact that CMS uses NLO cross sections from Prospino \cite{Beenakker:1996ed}
whereas we use leading order cross sections.
To convert cross section limits into limits on our parameter, $g_{\!_{DM}}$, we use a bisection method,
choosing the interval $g_{\!_{DM}}=0$ to $4\pi$. Larger values correspond to non-perturbative coupling, which would call into question the validity of the simplified model as an effective field theory.

We leave for future work consideration of other processes, such as the mono-jet process, 
$pp \rightarrow \chi \overline{\chi} j$ and associated production such
as $pp \rightarrow \chi \widetilde{u}$.  These processes lead to a distinct signature and
may help fill in the regime where the dark matter and the mediator are quasi-degenerate in mass, but
otherwise tend to provide weaker bounds than mediator pair production
for $g_{\!_{DM}} \lesssim g_s$.  It would be worthwhile to include these
processes in a full collider analysis.  In addition, the $q_L$ model could lead to leptonic signals due to
transitions between the $\widetilde{u}$ and $\widetilde{d}$ states resulting in $W$ bosons, once
one includes the mass splitting effects from the terms analogous to Equation~(\ref{eq:LFV}).

\begin{figure}
  \begin{subfigure}[b]{0.48\textwidth}
    \includegraphics[width=\textwidth]{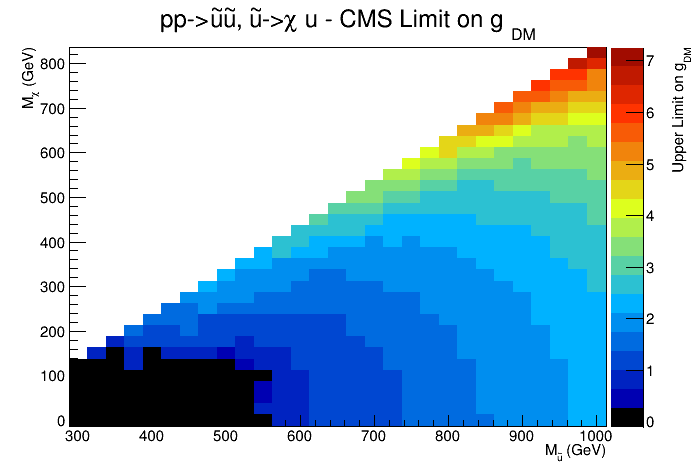}
    \caption{}
    \label{fig:CMS-a}
  \end{subfigure}

  \begin{subfigure}[b]{0.48\textwidth}
    \includegraphics[width=\textwidth]{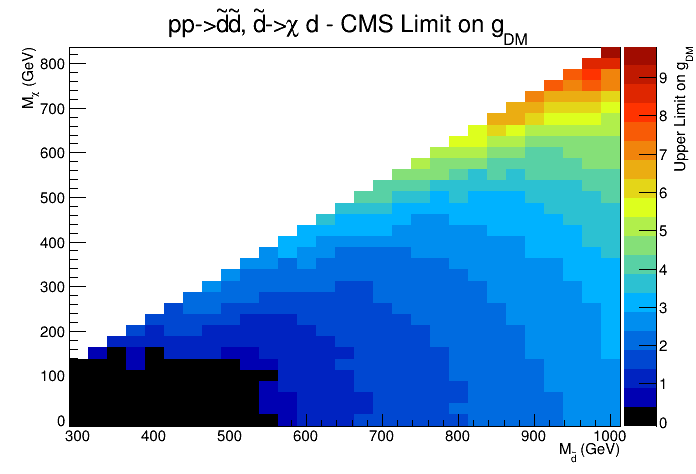}
    \caption{}
    \label{fig:CMS-b}
  \end{subfigure}

  \begin{subfigure}[b]{0.48\textwidth}
    \includegraphics[width=\textwidth]{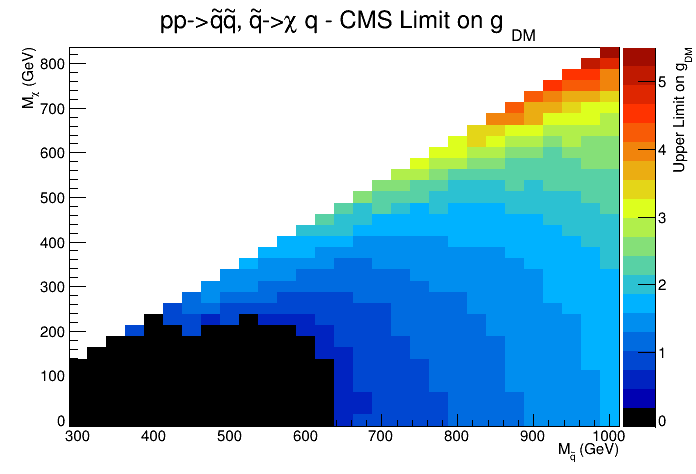}
    \caption{}
    \label{fig:CMS-c}
  \end{subfigure}
  \caption{Bounds on the the coupling $g_{\!_{DM}}$ for each
  of the three simplified
  models with Dirac Dark Matter, from the CMS collider bounds. (a) is the $u_R$ model, (b) the $d_R$ model, 
  and (c) is the $q_L$ model.}\label{fig:CMS}
\end{figure}

\begin{figure}
  \begin{subfigure}[b]{0.48\textwidth}
    \includegraphics[width=\textwidth]{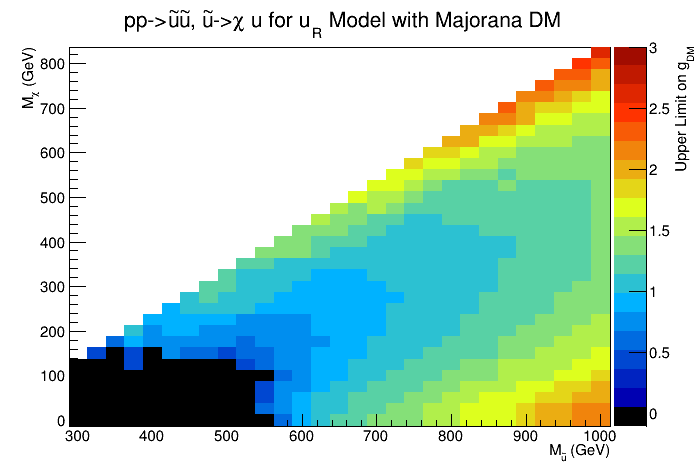}
    \caption{}
    \label{fig:CMSmaj-a}
  \end{subfigure}

  \begin{subfigure}[b]{0.48\textwidth}
    \includegraphics[width=\textwidth]{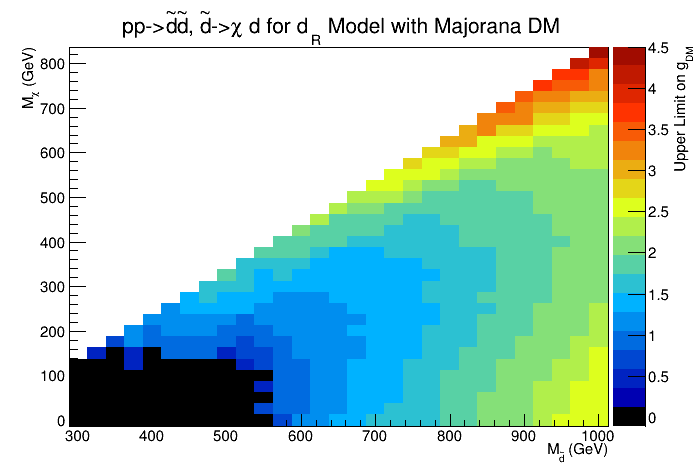}
    \caption{}
    \label{fig:CMSmaj-b}
  \end{subfigure}

  \begin{subfigure}[b]{0.48\textwidth}
    \includegraphics[width=\textwidth]{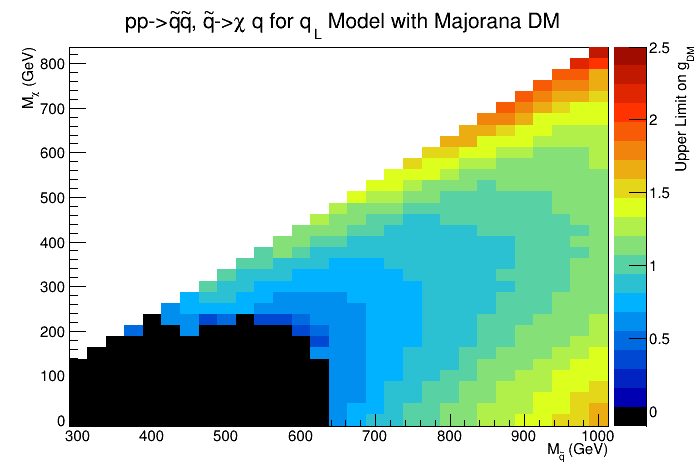}
    \caption{}
    \label{fig:CMSmaj-c}
  \end{subfigure}
  \caption{Bounds on the the coupling $g_{\!_{DM}}$ for all three models with Majorana Dark Matter, from the CMS collider bounds.}\label{fig:CMSmaj}
\end{figure}

The derived limits are shown for all three Dirac models in Figure~\ref{fig:CMS} and for the Majorana
models in Figure~\ref{fig:CMSmaj}.  In both figures, the white regions correspond to regions not considered
by CMS, because the mediator and dark matter are too degenerate in mass to efficiently pass the analysis
cuts (and in addition, much of this parameter space is theoretically inaccessible
because the scalar mediators are lighter than the $\chi$), whereas the black regions
are simply excluded by the CMS bounds for all values of $g_{\!_{DM}}$.  The excluded regions are largely
similar for both the Majorana and Dirac cases, where the limit is driven by gluon fusion into a
scalar mediator and its anti-particle.  However, in some regions the bounds are stronger for Majorana dark matter, 
which has the additional production process of two scalar mediators from a $qq$ initial state by exchanging the
Majorana $\chi$ in the $t$-channel.

\subsection{Direct Detection Bounds}

Elastic scattering of $\chi$ involves momentum transfers far below any mediator mass of interest, and the
contributions to spin-independent (SI) and spin-dependent (SD) scattering are most easily extracted
by computing the contact interaction between two $\chi$'s and two quarks.  Using the $u_R$ model as an
example, the matrix element for $\chi u \rightarrow \chi u$ is:
\begin{gather}
\mathcal{M} = (-ig_{\!_{DM}})^2 (\bar{\chi}P_Ru) \frac{i}{p^2-M_{\tilde{u}}^2} (\bar{u}P_L\chi)\\
\approx (-ig_{\!_{DM}})^2 (\bar{\chi}P_Ru) \frac{-i}{M^2_{\tilde{u}}-M^2_{\chi}} (\bar{u}P_L\chi)
\end{gather}
Where in the second line we take the limit of small momentum transfer. Applying a generalized
Fierz transformation \cite{Nieves:2003in} yields,
\begin{gather}
\mathcal{M} = \frac{ig_{\!_{DM}}^2}{M^2_{\tilde{u}}-M^2_{\chi}} \frac{1}{2} (\bar{\chi}\gamma^{\mu}P_L\chi)(\bar{u}\gamma_{\mu}P_Ru)\\
\begin{split}
=\frac{ig_{\!_{DM}}^2}{M^2_{\tilde{u}}-M^2_{\chi}} \frac{1}{8} [(\bar{\chi}\gamma^{\mu}\chi)(\bar{u}\gamma_{\mu}u) - (\bar{\chi}\gamma^{\mu}\gamma^5\chi)(\bar{u}\gamma_{\mu}\gamma_5u)\\ + (\bar{\chi}\gamma^{\mu}\gamma^5\chi)(\bar{u}\gamma_{\mu}u) - (\bar{\chi}\gamma^{\mu}\chi)(\bar{u}\gamma_{\mu}\gamma_5u)]
\end{split}\\
\approx \frac{ig_{\!_{DM}}^2}{M^2_{\tilde{u}}-M^2_{\chi}} \frac{1}{8} [(\bar{\chi}\gamma^{\mu}\chi)(\bar{u}\gamma_{\mu}u) - (\bar{\chi}\gamma^{\mu}\gamma^5\chi)(\bar{u}\gamma_{\mu}\gamma_5u)]
\end{gather}
where (as discussed in, e.g. \cite{Freytsis:2010ne}) we have dropped terms suppressed by
the dark matter velocity. The two remaining terms result in spin-independent and spin-dependent scattering,
respectively.
In the $u_R$ model, this results in cross sections for SI and SD scattering with a nucleon:
\begin{align}
\sigma_{SI}^{u_R} &= \frac{1}{64\pi} \frac{M_N^2M_{\chi}^2}{(M_N+M_{\chi})^2} \frac{g_{\!_{DM}}^4}{(M_{\tilde{u}}^2-M_{\chi}^2)^2} \left(1+\frac{Z}{A}\right)^2\\
\sigma_{SD}^{u_R} &= \frac{3}{64\pi} \frac{M_N^2M_{\chi}^2}{(M_N+M_{\chi})^2} \frac{g_{\!_{DM}}^4}{(M_{\tilde{u}}^2-M_{\chi}^2)^2} (\Delta u^N)^2
\end{align}
where $Z$, $A$, and 
$N=p, n$ specifies the nucleon of interest and the structure functions $\Delta u^N$ can be found,
for example, in Refs.~\cite{Freytsis:2010ne,Gondolo:2004sc}.  
Note that this theory has different SI cross sections for
protons and neutrons.

A similar calculation for the $d_R$ and $q_L$ Dirac models yields:
\begin{align}
\sigma_{SI}^{d_R} &= \frac{1}{64\pi} \frac{M_N^2M_{\chi}^2}{(M_N+M_{\chi})^2} \frac{g_{\!_{DM}}^4}{(M_{\tilde{d}}^2-M_{\chi}^2)^2} \left(2-\frac{Z}{A}\right)^2\\
\sigma_{SD}^{d_R} &= \frac{3}{64\pi} \frac{M_N^2M_{\chi}^2}{(M_N+M_{\chi})^2} \frac{g_{\!_{DM}}^4}{(M_{\tilde{d}}^2-M_{\chi}^2)^2} (\Delta d^N+\Delta s^N)^2\\
\sigma_{SI}^{q_L} &= \frac{9}{64\pi} \frac{M_N^2M_{\chi}^2}{(M_N+M_{\chi})^2} \frac{g_{\!_{DM}}^4}{(M_{\tilde{q}}^2-M_{\chi}^2)^2}\\
\sigma_{SD}^{q_L} &= \frac{3}{64\pi} \frac{M_N^2M_{\chi}^2}{(M_N+M_{\chi})^2} \frac{g_{\!_{DM}}^4}{(M_{\tilde{q}}^2-M_{\chi}^2)^2}
(\Delta u^N+\Delta d^N+\Delta s^N)^2
\end{align}
And likewise the cross sections for Majorana DM are also computed for each model:
\begin{align}
\sigma_{SD}^{u_R} &= \frac{3}{16\pi} \frac{M_N^2M_{\chi}^2}{(M_N+M_{\chi})^2} \frac{g_{\!_{DM}}^4}{(M_{\tilde{d}}^2-M_{\chi}^2)^2} (\Delta u^N)^2\\
\sigma_{SD}^{d_R} &= \frac{3}{16\pi} \frac{M_N^2M_{\chi}^2}{(M_N+M_{\chi})^2} \frac{g_{\!_{DM}}^4}{(M_{\tilde{d}}^2-M_{\chi}^2)^2} (\Delta d^N+\Delta s^N)^2\\
\sigma_{SD}^{q_L} &= \frac{3}{16\pi} \frac{M_N^2M_{\chi}^2}{(M_N+M_{\chi})^2} \frac{g_{\!_{DM}}^4}{(M_{\tilde{d}}^2-M_{\chi}^2)^2} (\Delta u^N+\Delta d^N+\Delta s^N)^2
\end{align}
Note that since a Majorana fermion has a vanishing vector bilinear, 
there are only spin-dependent cross-sections for the Majorana DM cases\footnote{It would be interesting
to compute the induced SI cross section at one-loop for this class of simplified model.}.

For Dirac dark matter, this class of simplified models results in roughly comparable spin-independent and dependent cross sections.  Since the SI bounds at a given dark matter mass
are currently much lower than the corresponding SD ones, they dominate
the constraints.  We translate the limits from
XENON100 \cite{Aprile:2012nq} and XENON10 \cite{Angle:2011th} into bounds on $g_{\!_{DM}}$
at every point in the plane of the dark matter and mediator masses.  For Majorana dark matter,
the dominant constraint is from the SD interaction, with the best experimental limits for
neutrons and protons arising from XENON100 \cite{Aprile:2013doa} and
PICASSO \cite{Archambault:2012pm}, respectively.  It turned out in all cases that the PICASSO limit was
weaker than the CMS limit corresponding to that point of parameter space.

\begin{figure}
  \begin{subfigure}[b]{0.46\textwidth}
    \includegraphics[width=\textwidth]{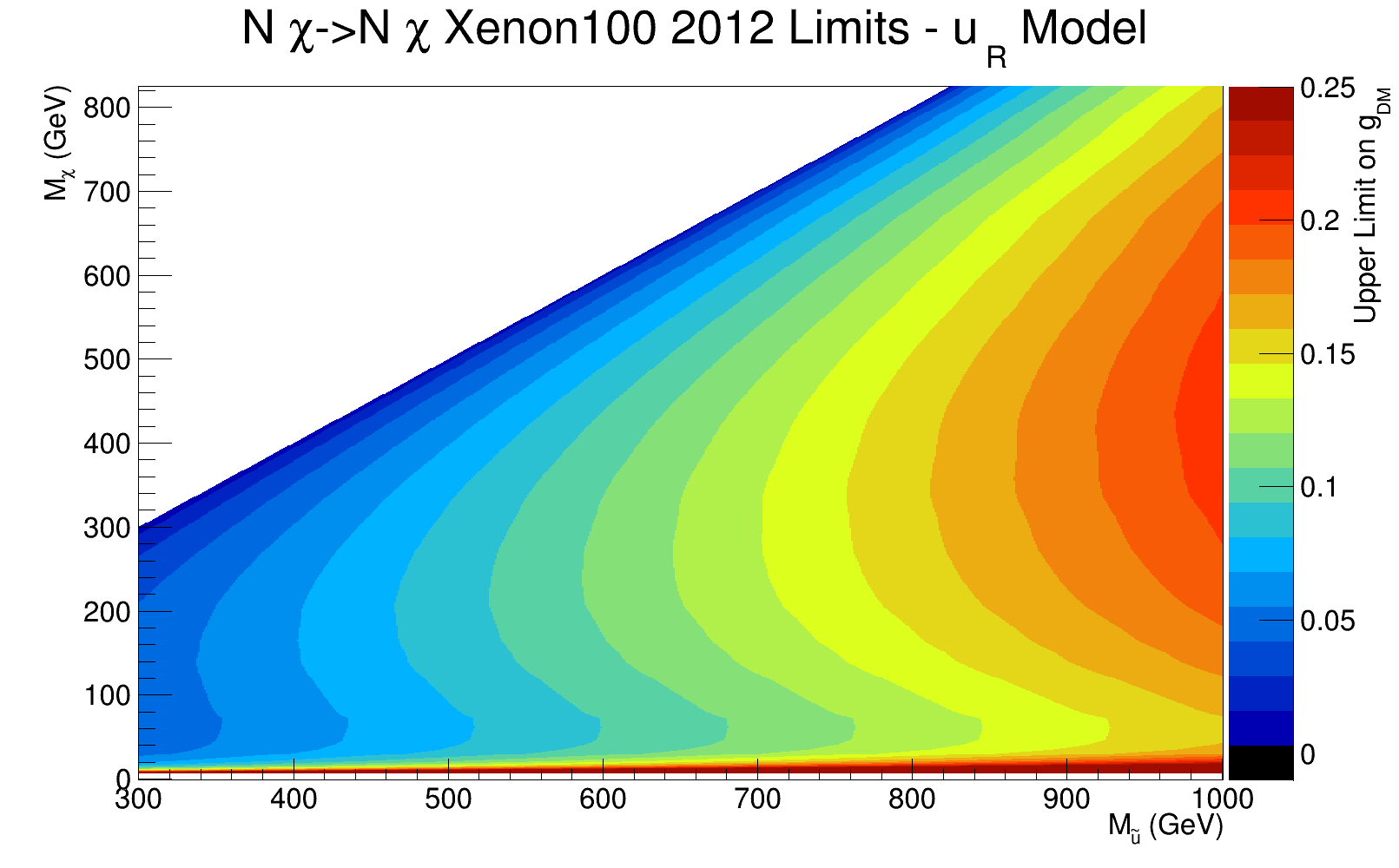}
    \caption{}
    \label{fig:Xe100-a}
  \end{subfigure}

  \begin{subfigure}[b]{0.46\textwidth}
    \includegraphics[width=\textwidth]{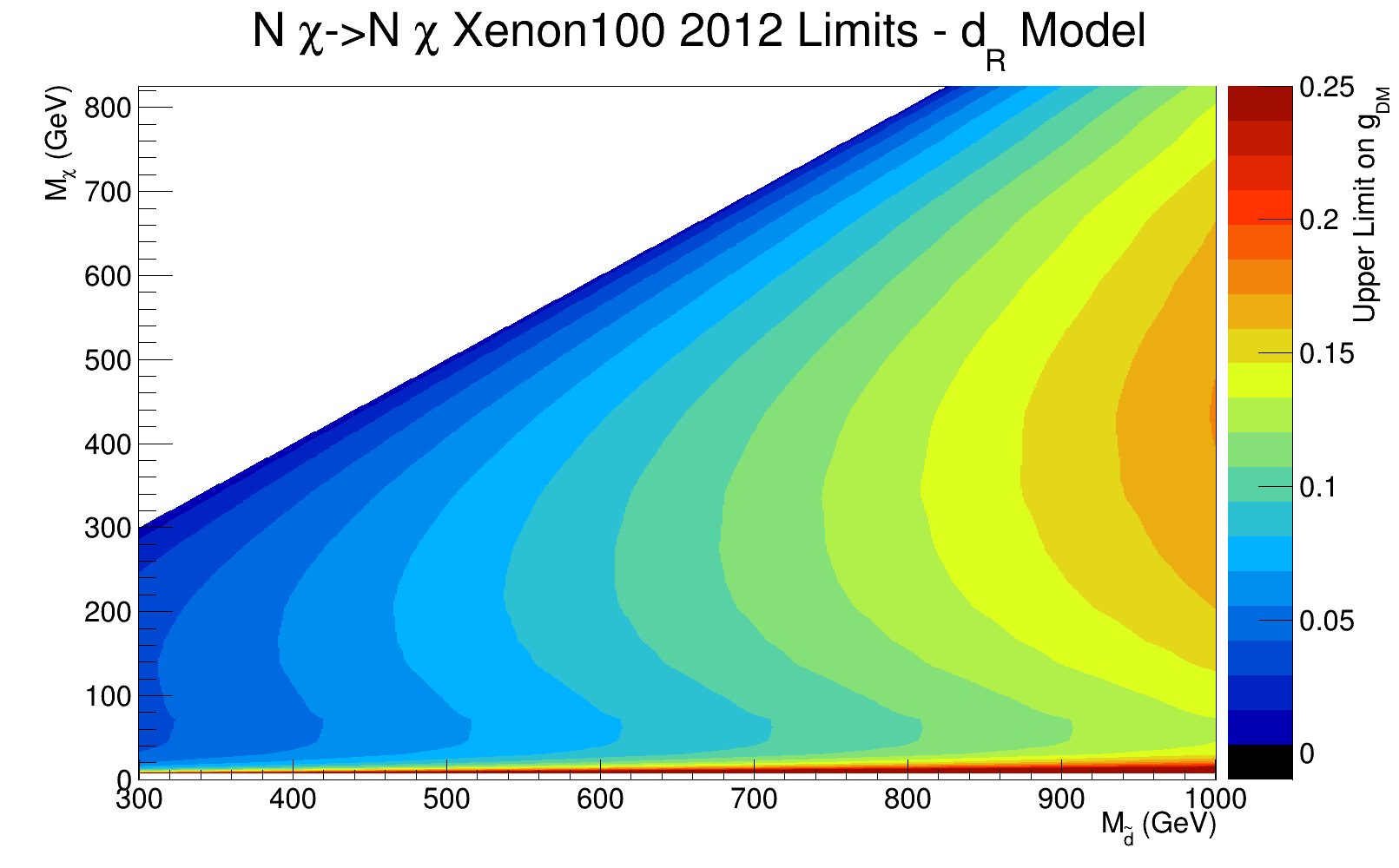}
    \caption{}
    \label{fig:Xe100-b}
  \end{subfigure}

  \begin{subfigure}[b]{0.46\textwidth}
    \includegraphics[width=\textwidth]{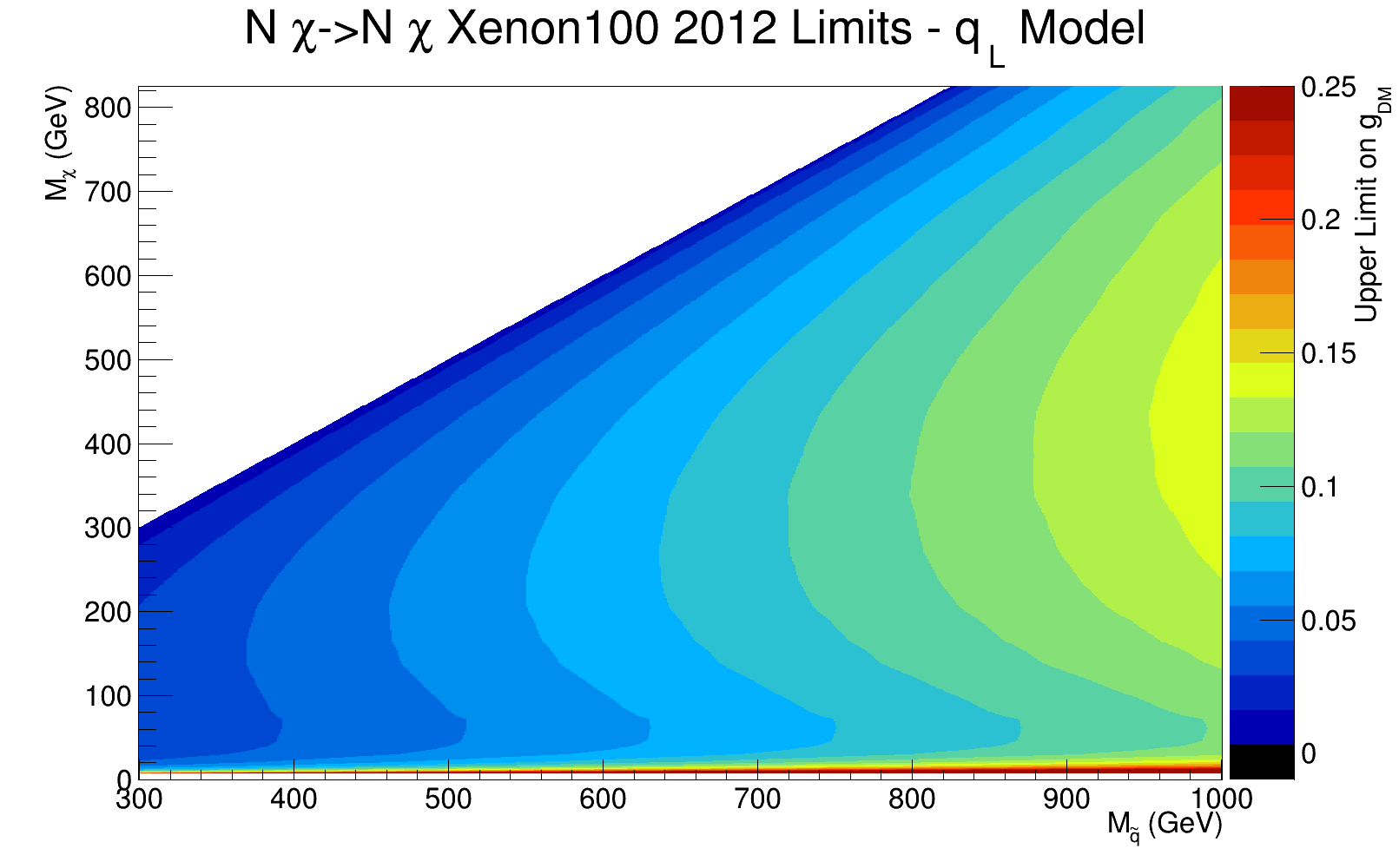}
    \caption{}
    \label{fig:Xe100-c}
  \end{subfigure}
  \caption{Bounds on the the coupling $g_{\!_{DM}}$ for all three models with Dirac Dark Matter, from the spin-independent XENON100 Limits.}\label{fig:Xe100}
\end{figure}

\begin{figure}
  \begin{subfigure}[b]{0.46\textwidth}
    \includegraphics[width=\textwidth]{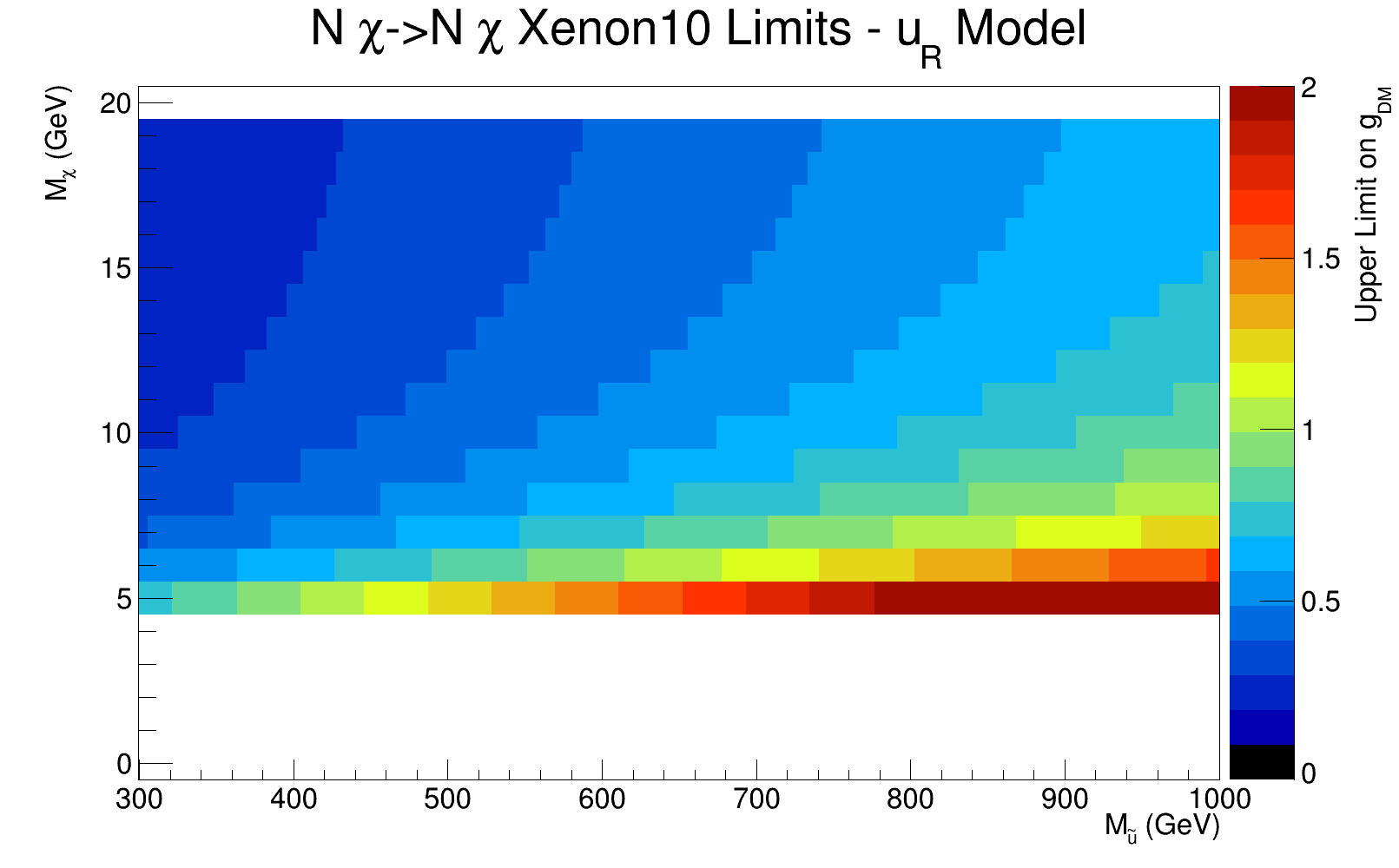}
    \caption{}
    \label{fig:Xe10-a}
  \end{subfigure}

  \begin{subfigure}[b]{0.46\textwidth}
    \includegraphics[width=\textwidth]{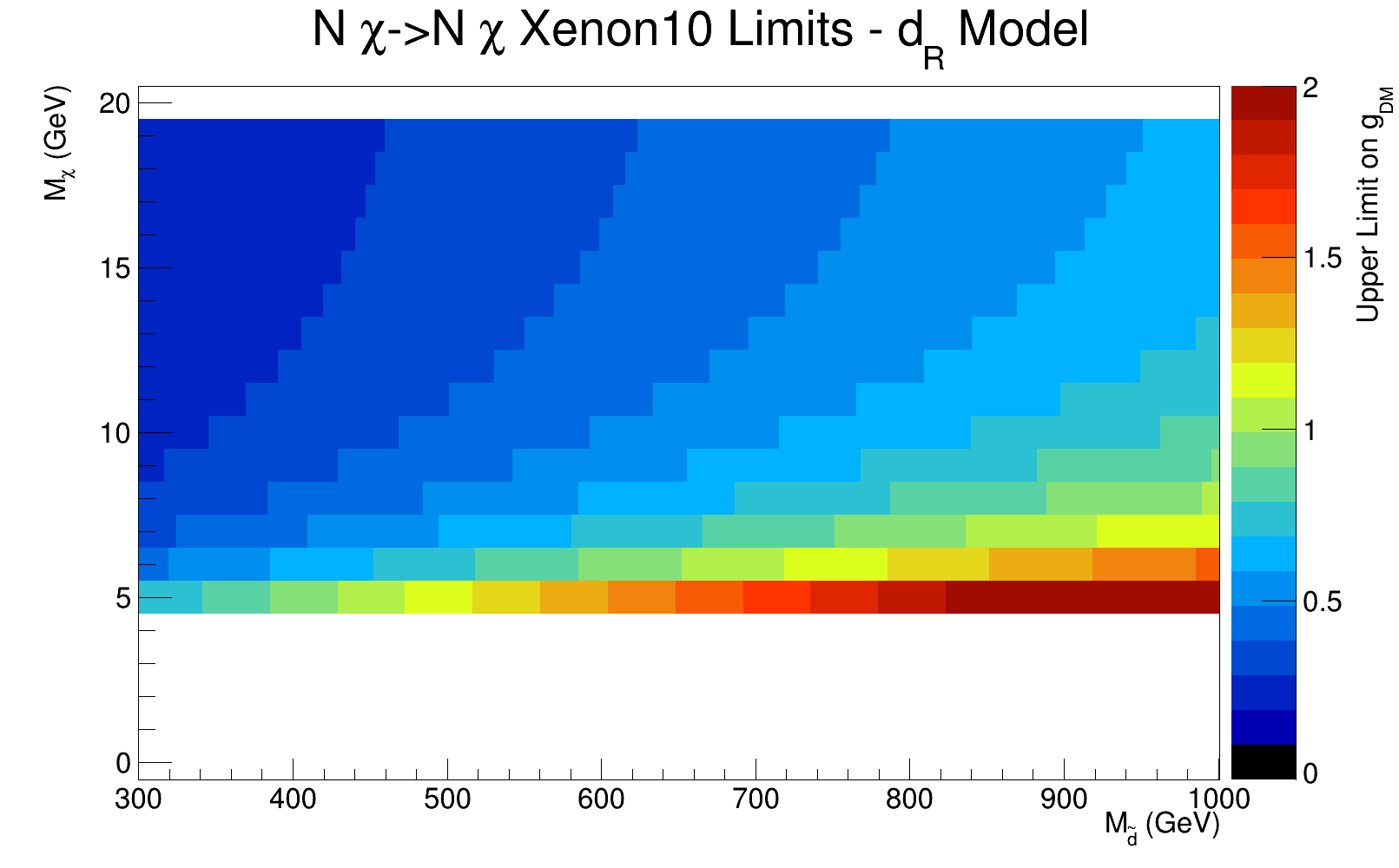}
    \caption{}
    \label{fig:Xe10-b}
  \end{subfigure}

  \begin{subfigure}[b]{0.46\textwidth}
    \includegraphics[width=\textwidth]{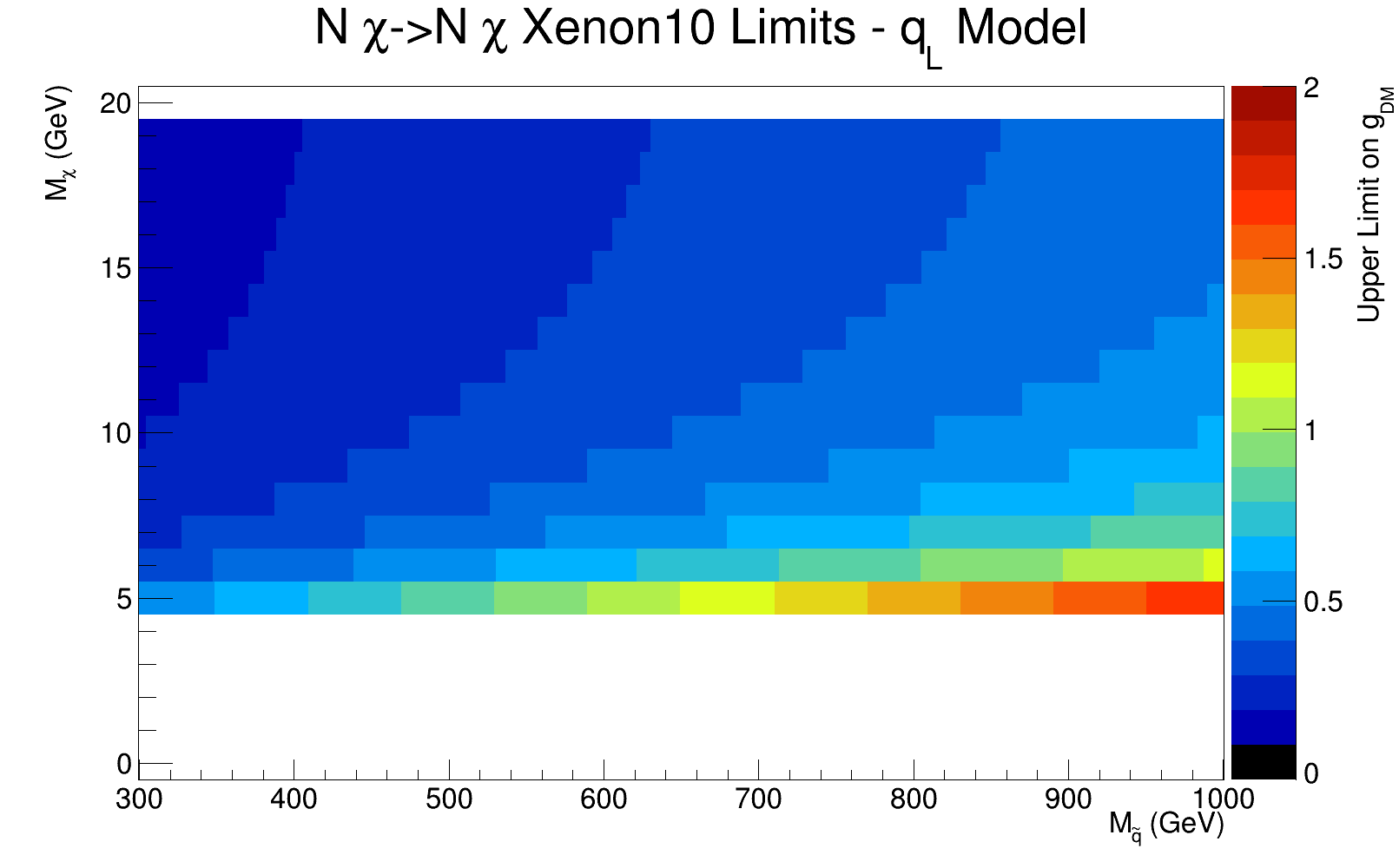}
    \caption{}
    \label{fig:Xe10-c}
  \end{subfigure}
  \caption{Dirac Dark Matter bounds on $g_{\!_{DM}}$ from the spin-indepedent XENON10 Limits.}\label{fig:Xe10}
\end{figure}

\begin{figure}
  \begin{subfigure}[b]{0.46\textwidth}
    \includegraphics[width=\textwidth]{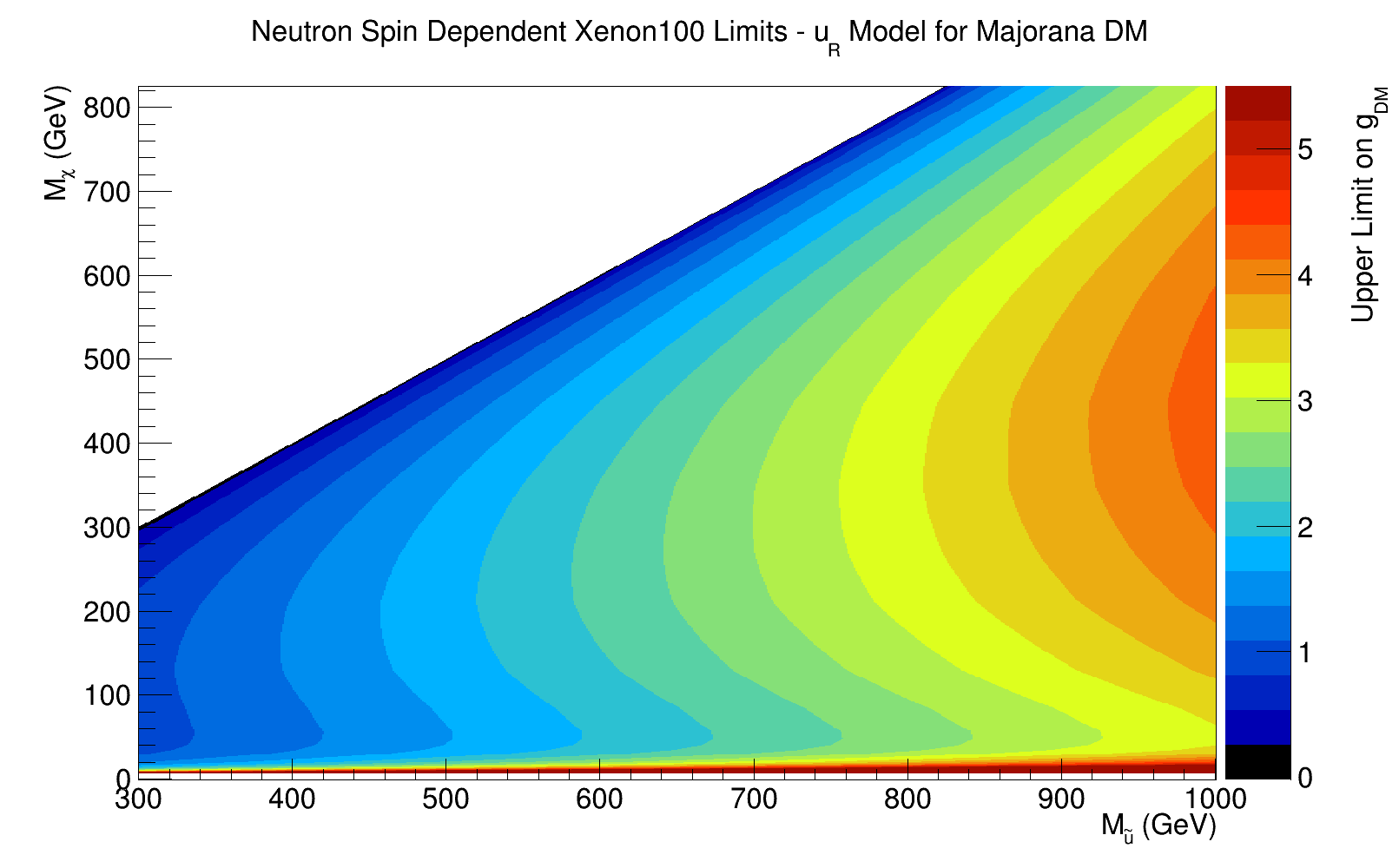}
    \caption{}
    \label{fig:sdXe-a}
  \end{subfigure}

  \begin{subfigure}[b]{0.46\textwidth}
    \includegraphics[width=\textwidth]{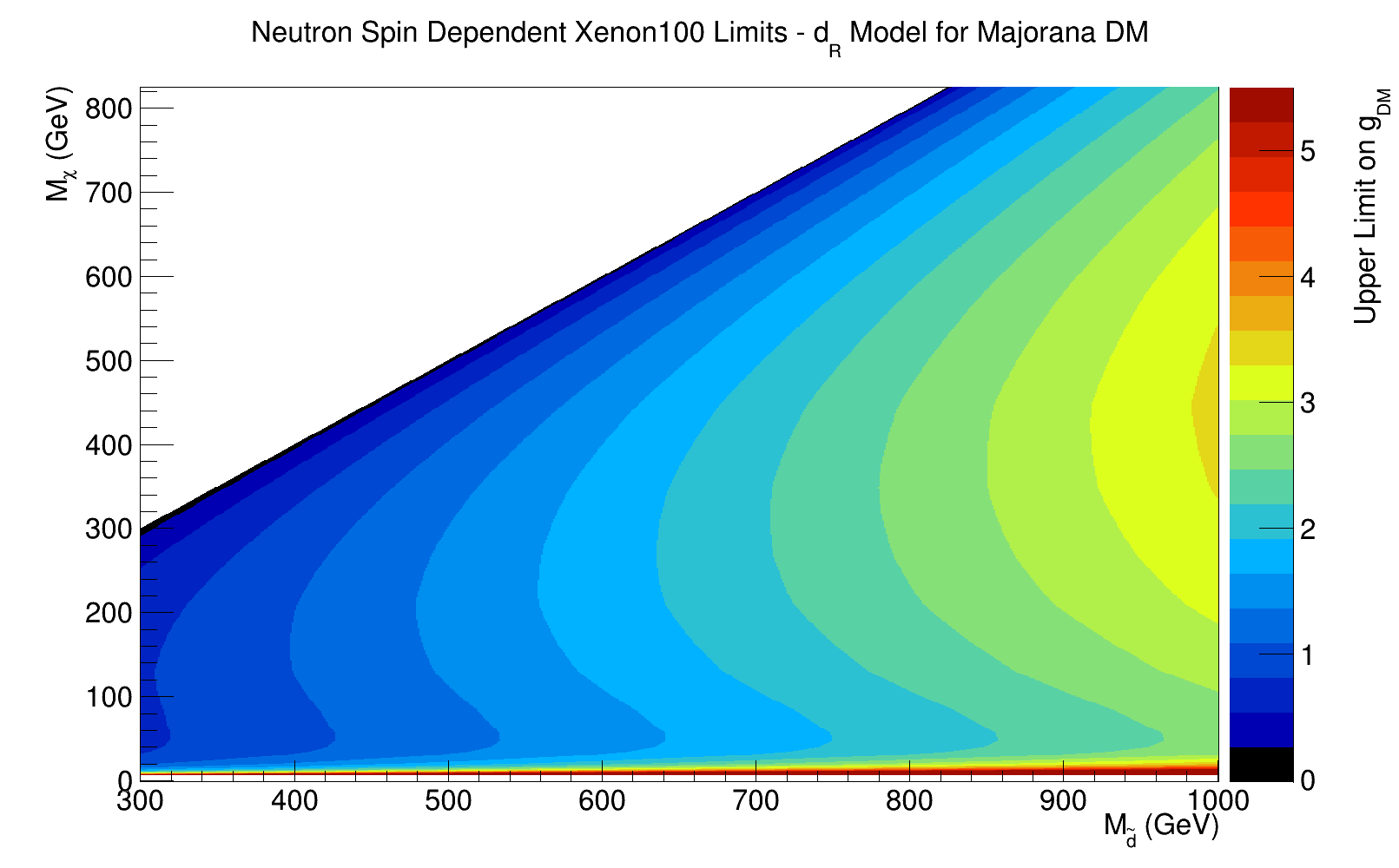}
    \caption{}
    \label{fig:sdXe-b}
  \end{subfigure}

  \begin{subfigure}[b]{0.46\textwidth}
    \includegraphics[width=\textwidth]{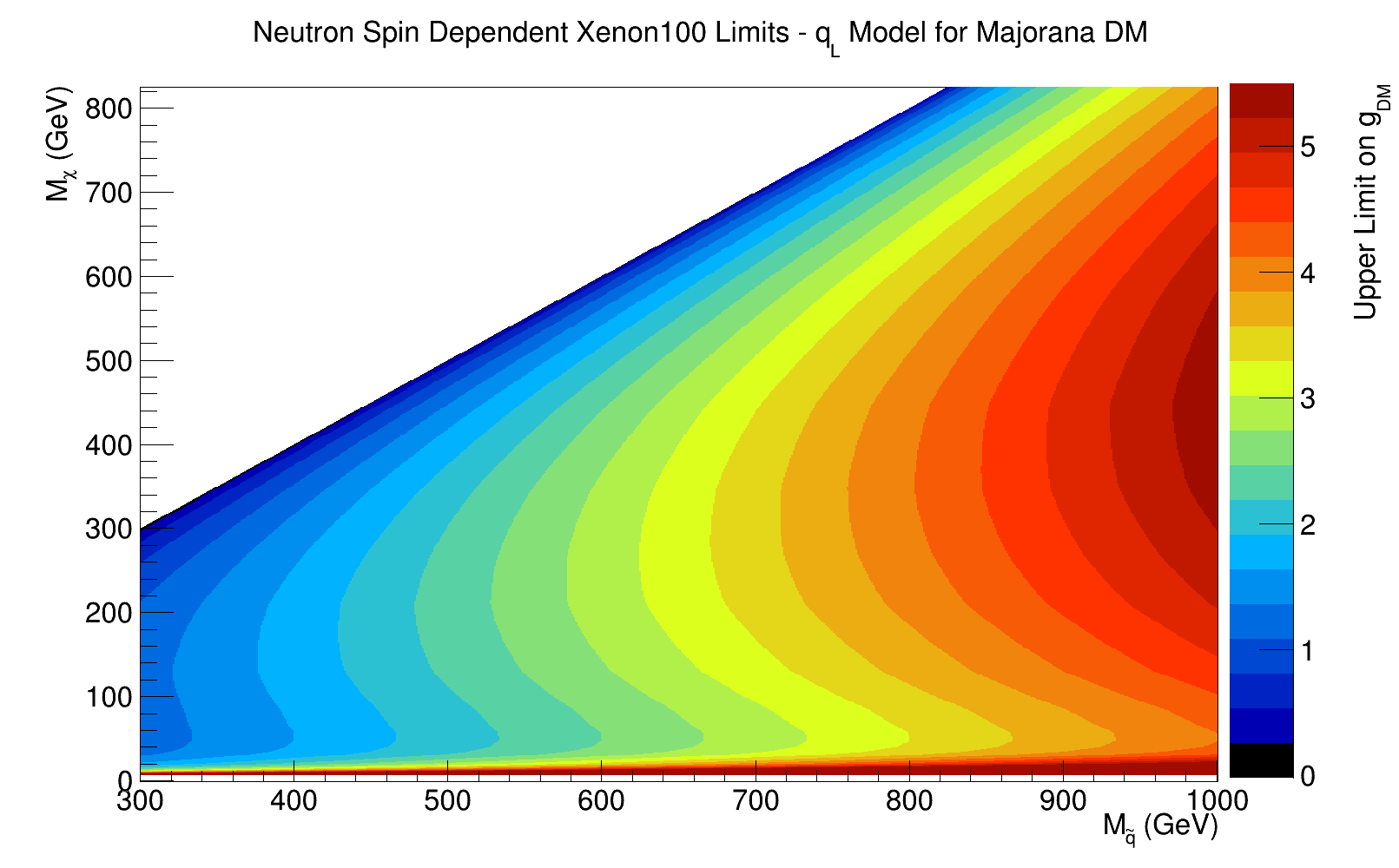}
    \caption{}
    \label{fig:sdXe-c}
  \end{subfigure}
  \caption{Bounds on $g_{\!_{DM}}$ from neutron-WIMP spin-dependent XENON100 Limits on Majorana Dark Matter.}\label{fig:sdXe}
\end{figure}

Applying these limits, the results for Dirac dark matter are shown in
Figure~\ref{fig:Xe100} (with a zoomed in view of the small dark matter mass region in Figure~\ref{fig:Xe10} from Xenon10 limits), 
and for Majorana dark matter in \ref{fig:sdXe}.  In these plots, the white regions indicate
theoretically inaccessible regions with colored mediators lighter than the dark matter itself.
For large $M_{\chi}$, the XENON limits are slowly varying and nearly linear. In this regime, the elastic scattering 
cross section is approximately independent of $M_{\chi}$ and we observe the expected behavior where the 
curves of constant $g_{\!_{DM}}$ go like $(M_{\chi}^2-M_{\tilde{q}}^2)^{-2}$, similar to the dependence of the direct detection cross section 
on the two masses. For $M_{\chi} \lesssim 10$~GeV, the limit 
on $g_{\!_{DM}}$ from XENON weakens very rapidly as the experiments approach their threshold for 
observable scattering. In turn, this allows $g_{\!_{DM}}$ to be much larger as $M_{\chi}$ decreases.
This feature is illustrated in \ref{fig:Xe10} where we have zoomed into a much smaller scale of 
$M_{\chi}$ to more clearly show the behavior in this regime.

\section{Results}

\begin{figure}
  \begin{subfigure}[b]{0.46\textwidth}
    \includegraphics[width=\textwidth]{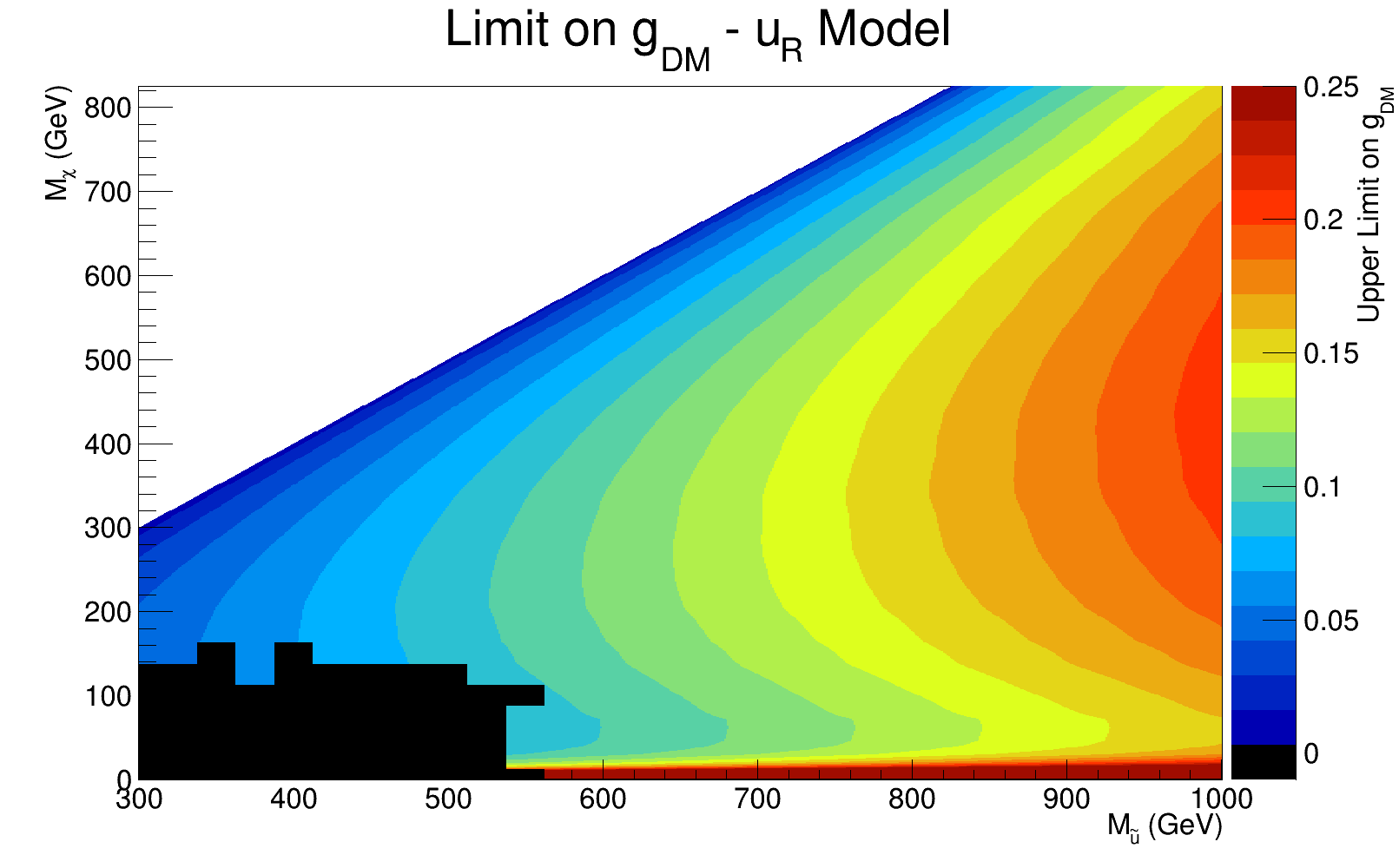}
    \caption{}
    \label{fig:Dbound-a}
  \end{subfigure}

  \begin{subfigure}[b]{0.46\textwidth}
    \includegraphics[width=\textwidth]{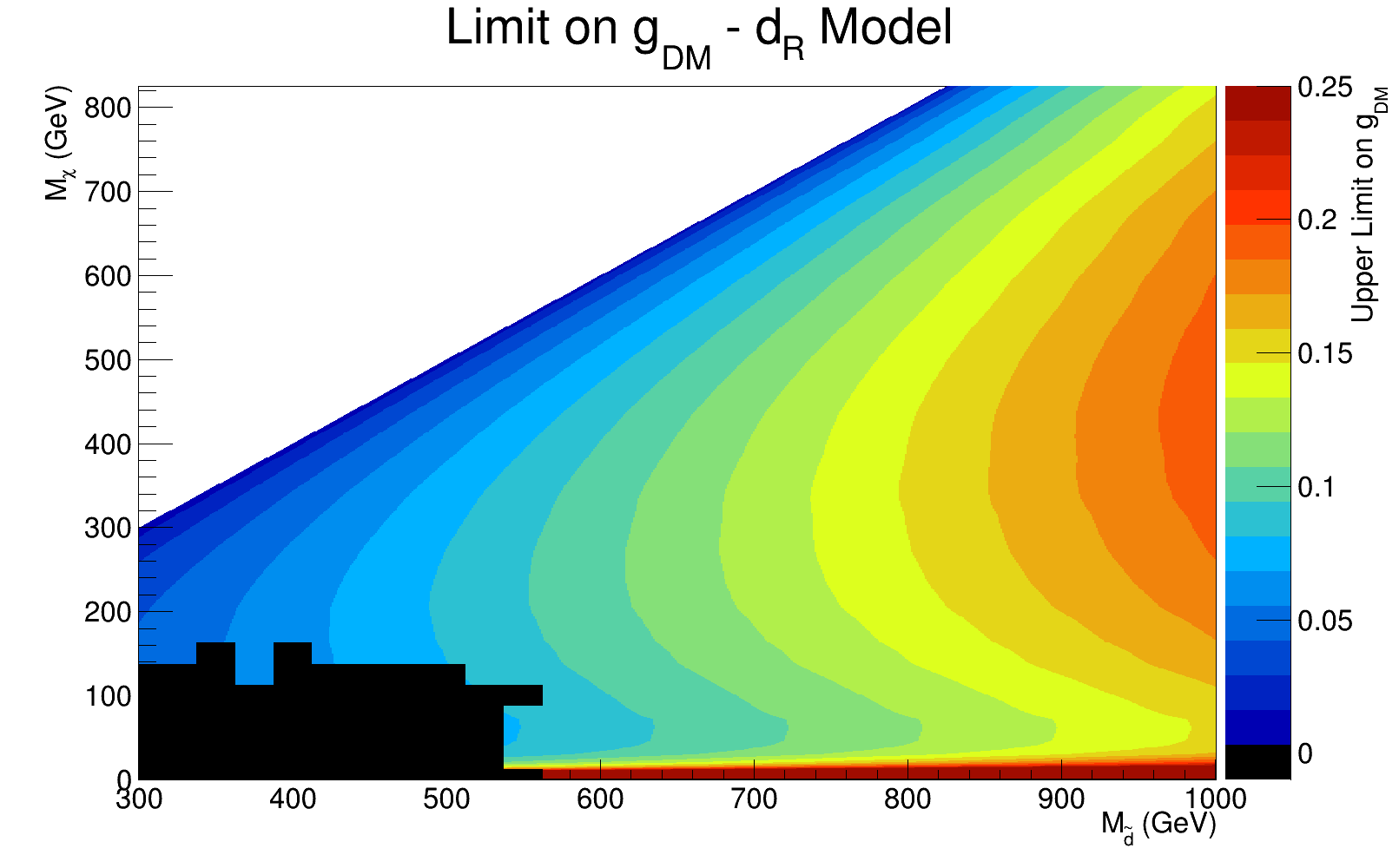}
    \caption{}
    \label{fig:Dbound-b}
  \end{subfigure}

  \begin{subfigure}[b]{0.46\textwidth}
    \includegraphics[width=\textwidth]{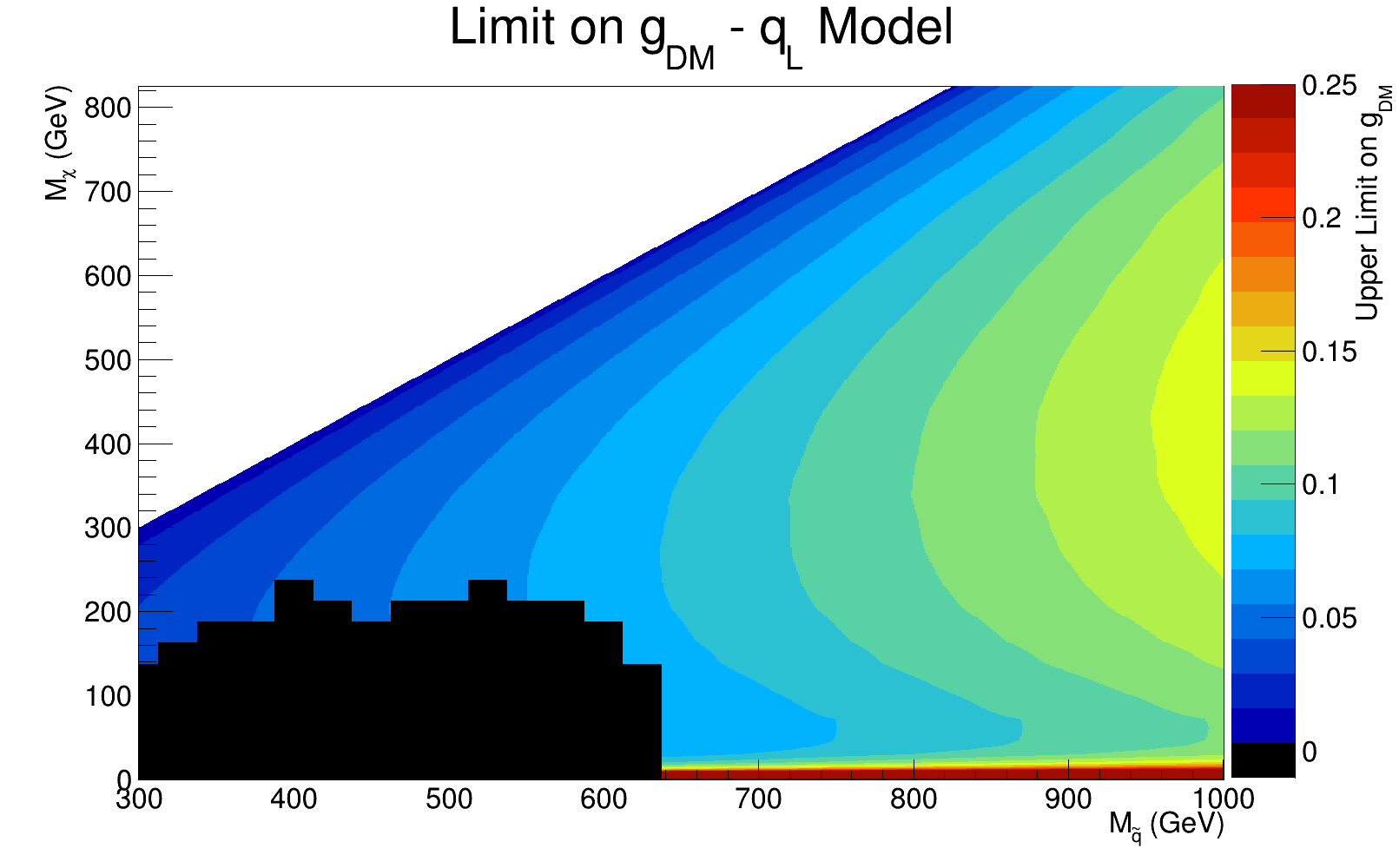}
    \caption{}
    \label{fig:Dbound-c}
  \end{subfigure}
  \caption{The combined lowest bounds on $g_{\!_{DM}}$ from CMS, XENON100, and XENON10 for Dirac Dark Matter.}\label{fig:Dbound}
\end{figure}

\begin{figure}
  \begin{subfigure}[b]{0.46\textwidth}
    \includegraphics[width=\textwidth]{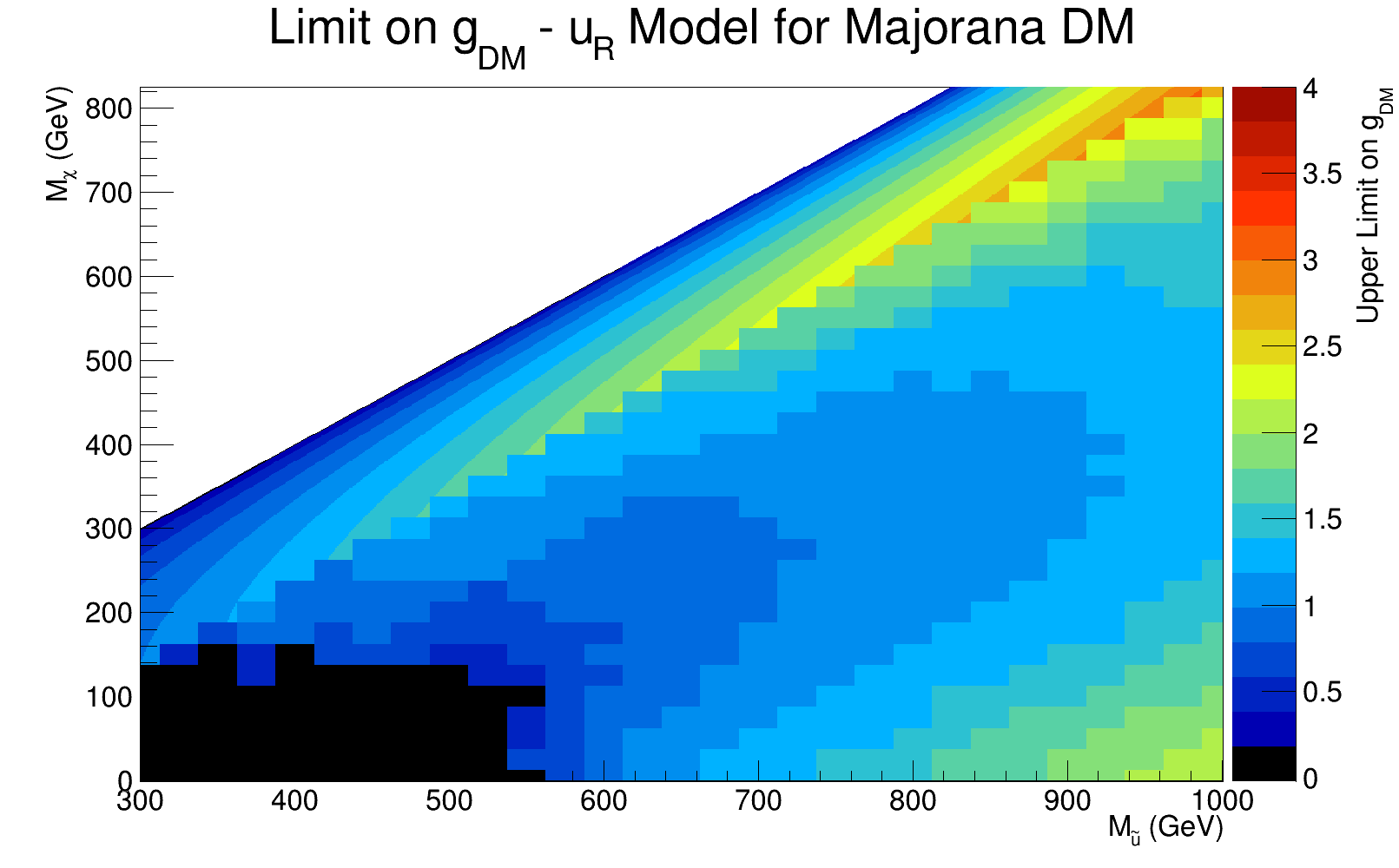}
    \caption{}
    \label{fig:Mbound-a}
  \end{subfigure}

  \begin{subfigure}[b]{0.46\textwidth}
    \includegraphics[width=\textwidth]{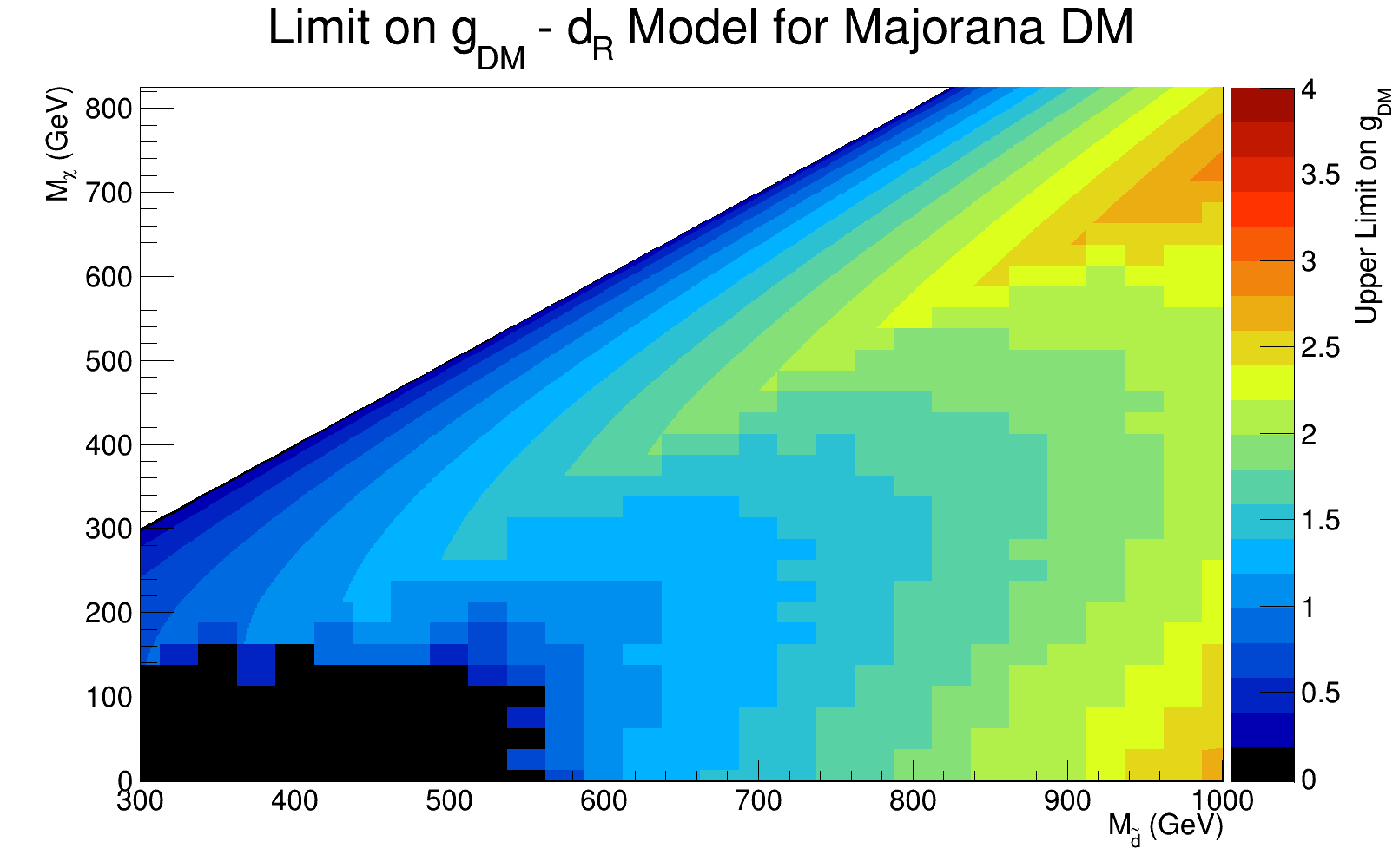}
    \caption{}
    \label{fig:Mbound-b}
  \end{subfigure}

  \begin{subfigure}[b]{0.46\textwidth}
    \includegraphics[width=\textwidth]{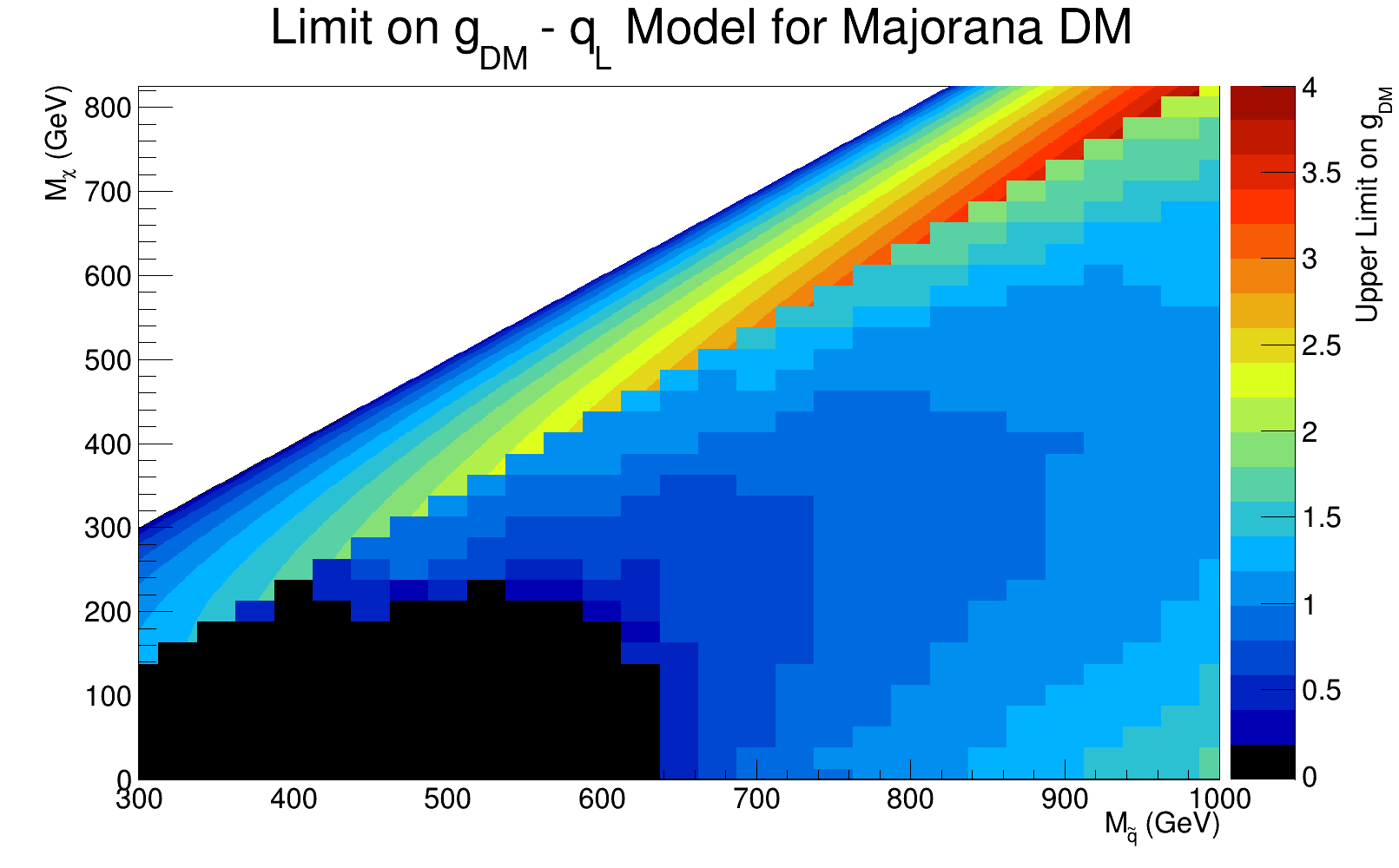}
    \caption{}
    \label{fig:Mbound-c}
  \end{subfigure}

  \caption{The combined lowest limit on $g_{\!_{DM}}$ from CMS and XENON100 for Majorana Dark Matter.}\label{fig:Mbound}
\end{figure}

We determine combined limits by first applying the CMS bounds, and overlaying those
from direct detection, replacing the existing limit on $g_{\!_{DM}}$ only when the direct detection limit is 
stronger than the collider one.  The resulting limits are shown in Figure~\ref{fig:Dbound} for Dirac
dark matter, and in Figure~\ref{fig:Mbound} for Majorana dark matter.
Generically, the most tightly bounded case is the $q_L$ model, due to its larger number of colored states
which can either be produced at the collider or mediate elastic scattering.  However, 
the contribution to the SD elastic scattering in the Majorana model exhibits destructive interference between
up-quarks and down-quarks, which allows for weaker constraints in that case.

These results illustrate an important complementarity between experimental probes of dark matter.  In addition
to the well-known cases where colliders dominate direct detection bounds for very light dark matter and/or
dark matter with predominantly spin-dependent interactions, we also see a different kind of complementarity.
Because colliders can access the mediators directly through QCD production, 
they can rule out cases where the mediator mass is light enough.  Instead, direct detection cannot rule out
a model, but generically places the strongest limits on $g_{\!_{DM}}$ for Dirac dark matter
which has SI interactions.  For the Majorana cases, the SD limits are typically weak enough so as to be
subdominant to the collider bounds, except for very degenerate spectra.

\subsection{Forecasts for Future Experiments}

\begin{figure}
  \begin{subfigure}[b]{0.46\textwidth}
    \includegraphics[width=\textwidth]{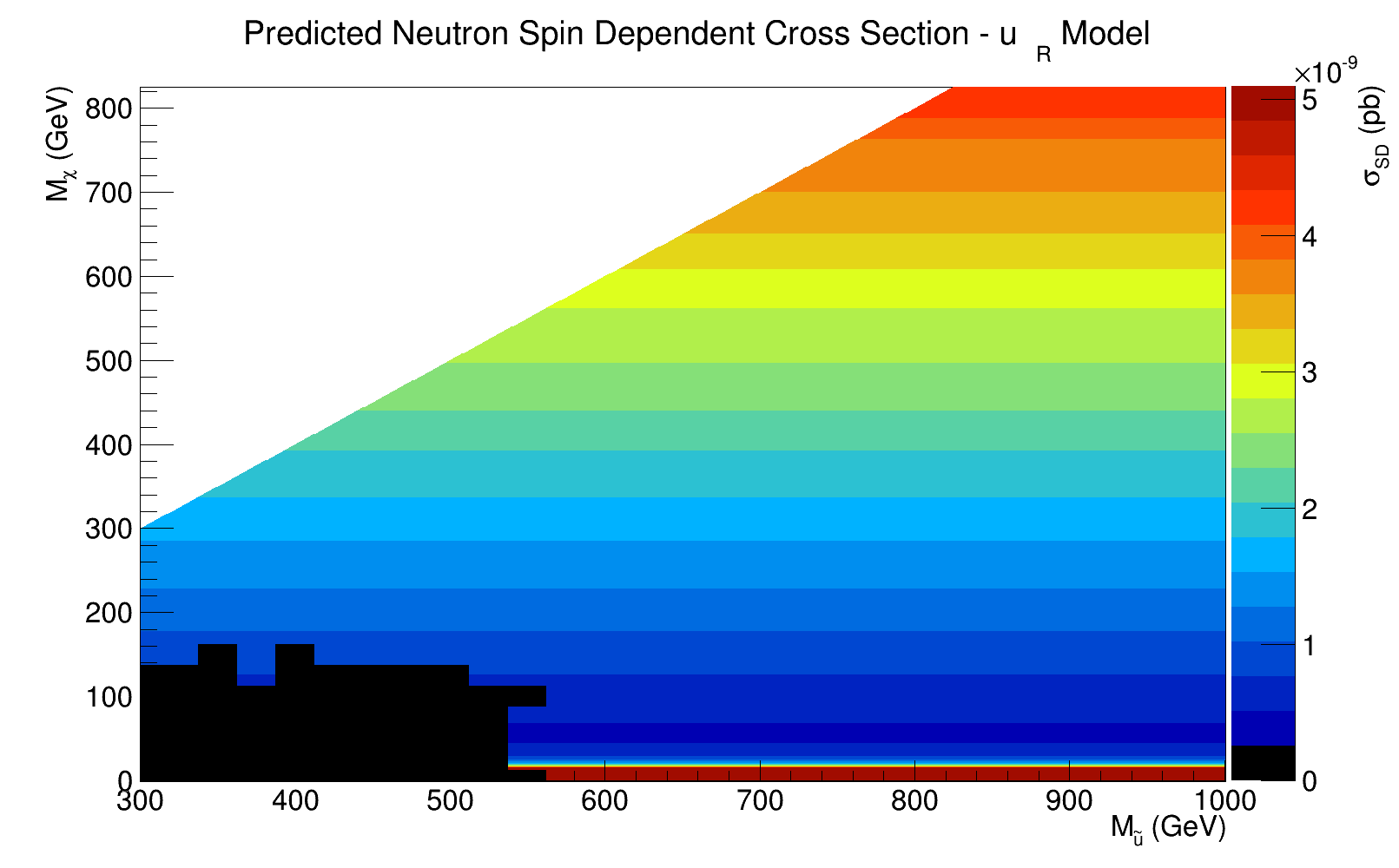}
    \caption{}
    \label{fig:sdpredn-a}
  \end{subfigure}

  \begin{subfigure}[b]{0.46\textwidth}
    \includegraphics[width=\textwidth]{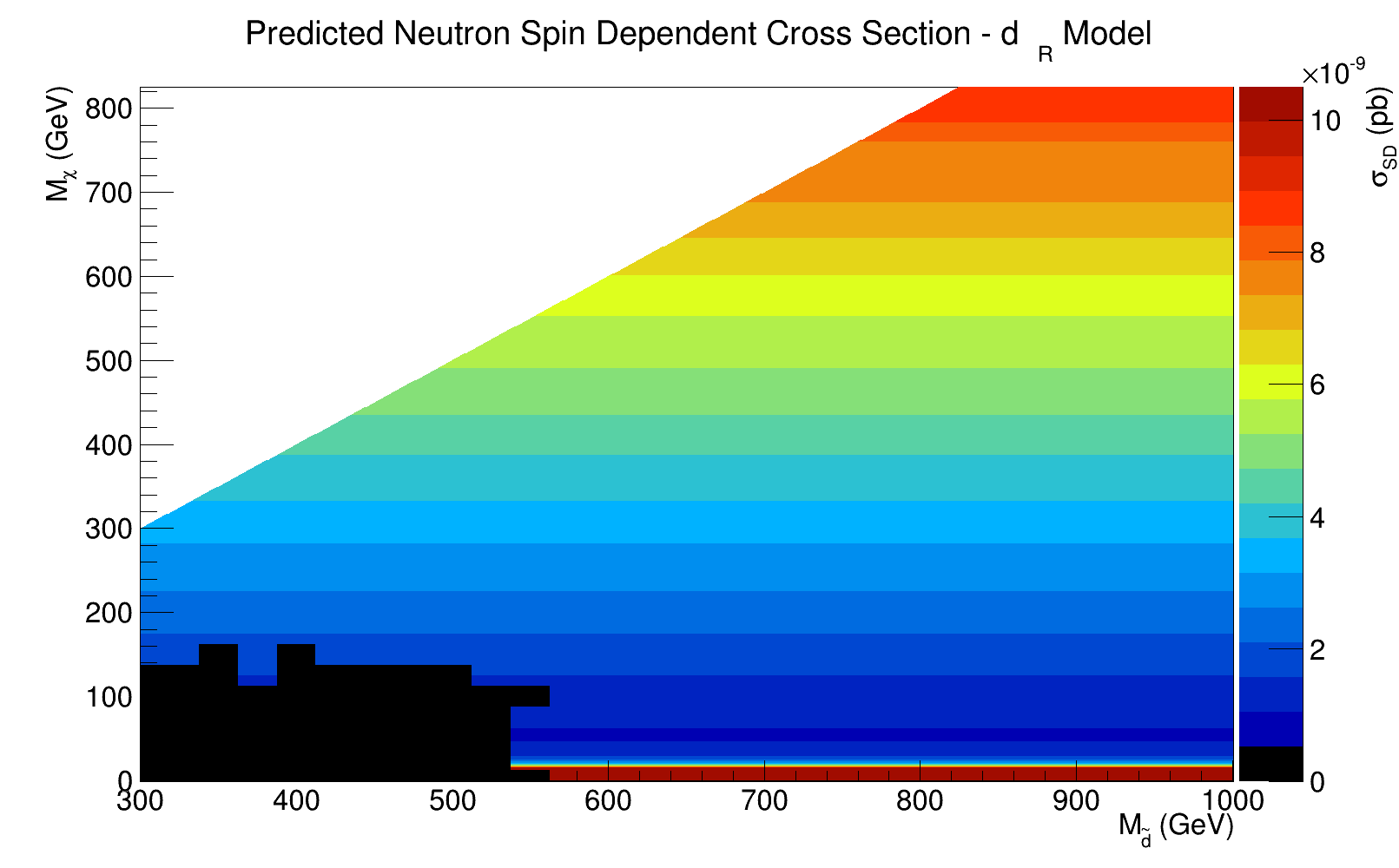}
    \caption{}
    \label{fig:sdpredn-b}
  \end{subfigure}

  \begin{subfigure}[b]{0.46\textwidth}
    \includegraphics[width=\textwidth]{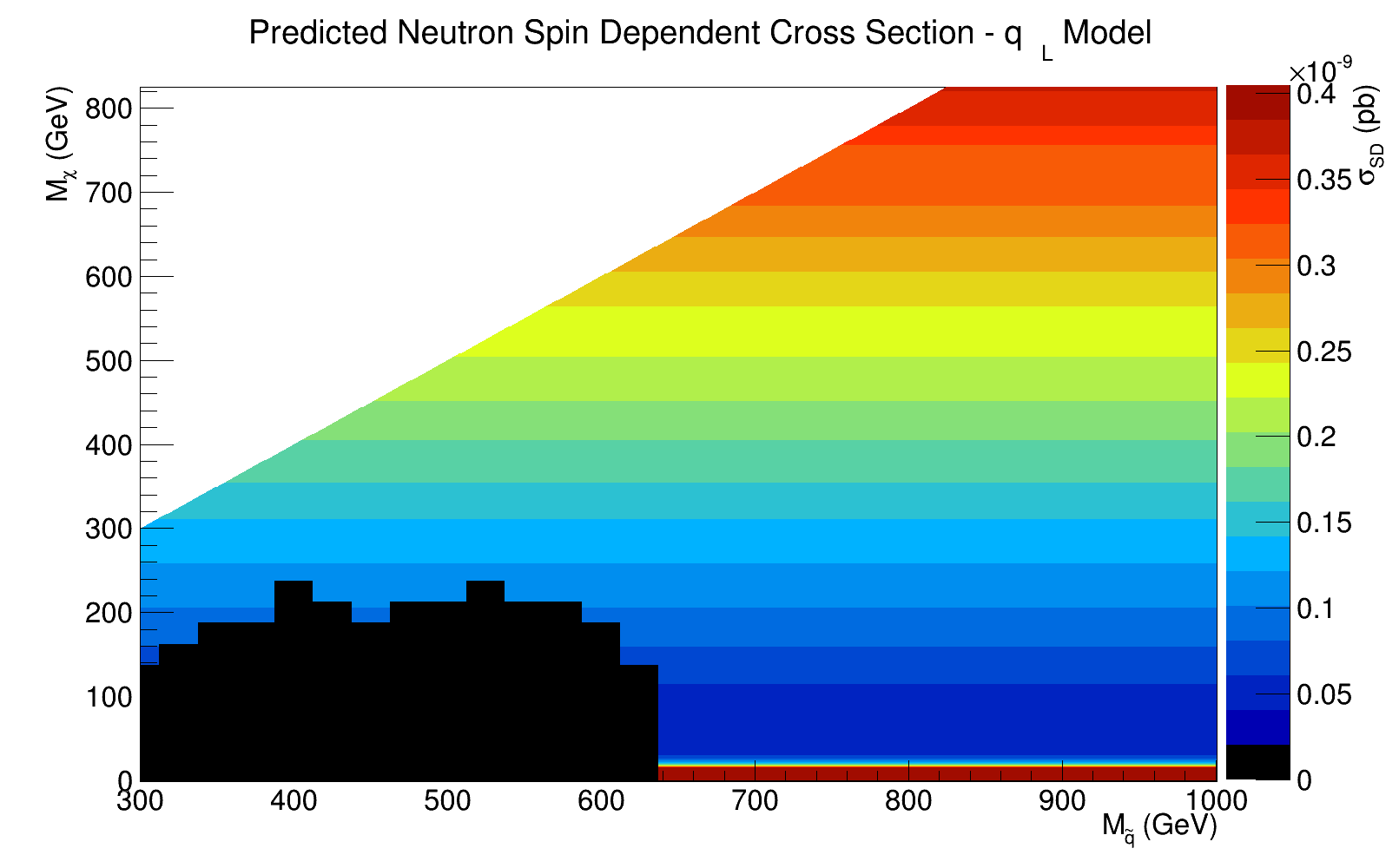}
    \caption{}
    \label{fig:sdpredn-c}
  \end{subfigure}
  \caption{The predicted maximum spin-dependent neutron-DM cross section from the combined Collider and Direct Detection bounds for Dirac Dark Matter.}\label{fig:sdpredn}
\end{figure}

\begin{figure}
  \begin{subfigure}[b]{0.46\textwidth}
    \includegraphics[width=\textwidth]{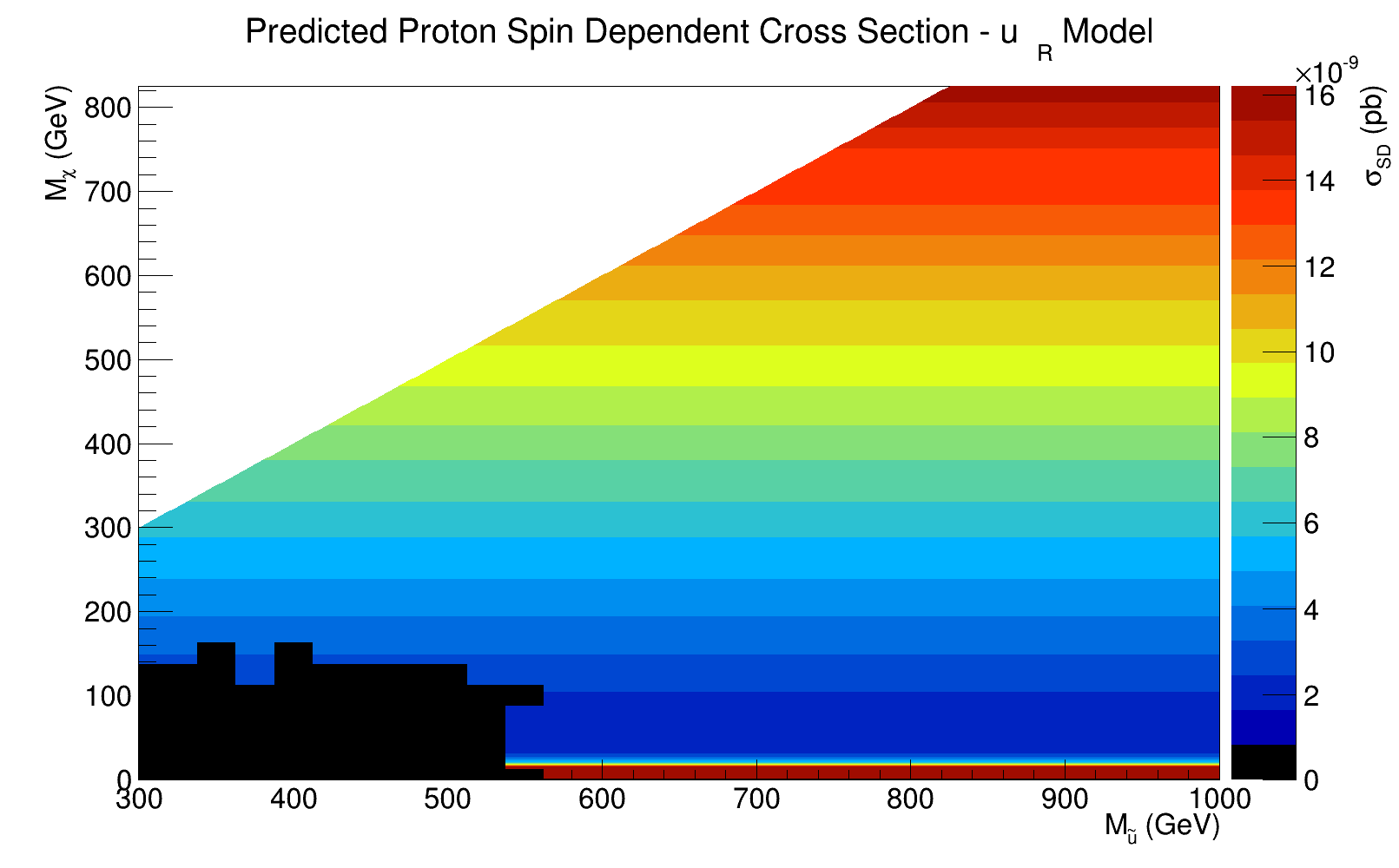}
    \caption{}
    \label{fig:sdpredp-a}
  \end{subfigure}

  \begin{subfigure}[b]{0.46\textwidth}
    \includegraphics[width=\textwidth]{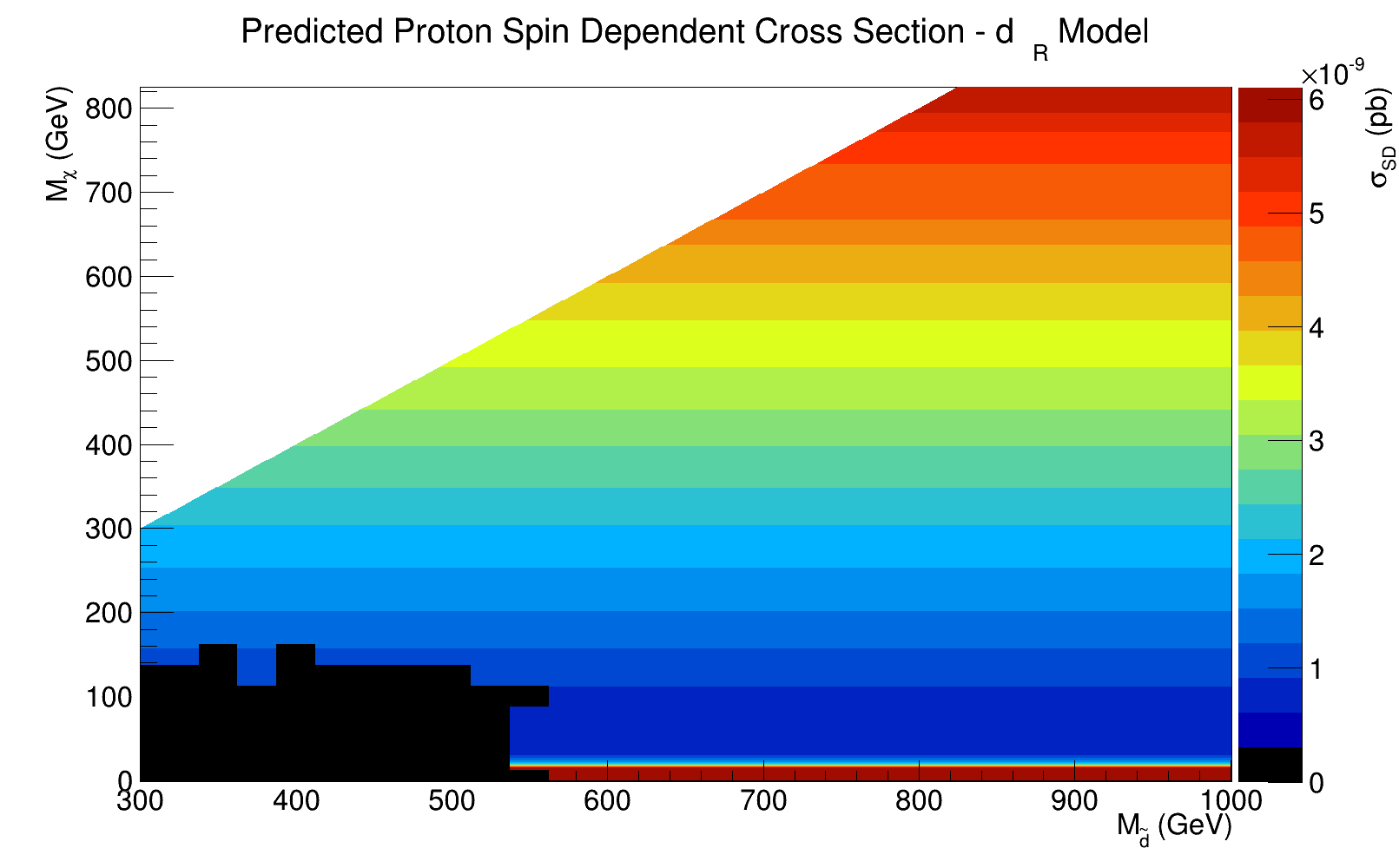}
    \caption{}
    \label{fig:sdpredp-b}
  \end{subfigure}

  \begin{subfigure}[b]{0.46\textwidth}
    \includegraphics[width=\textwidth]{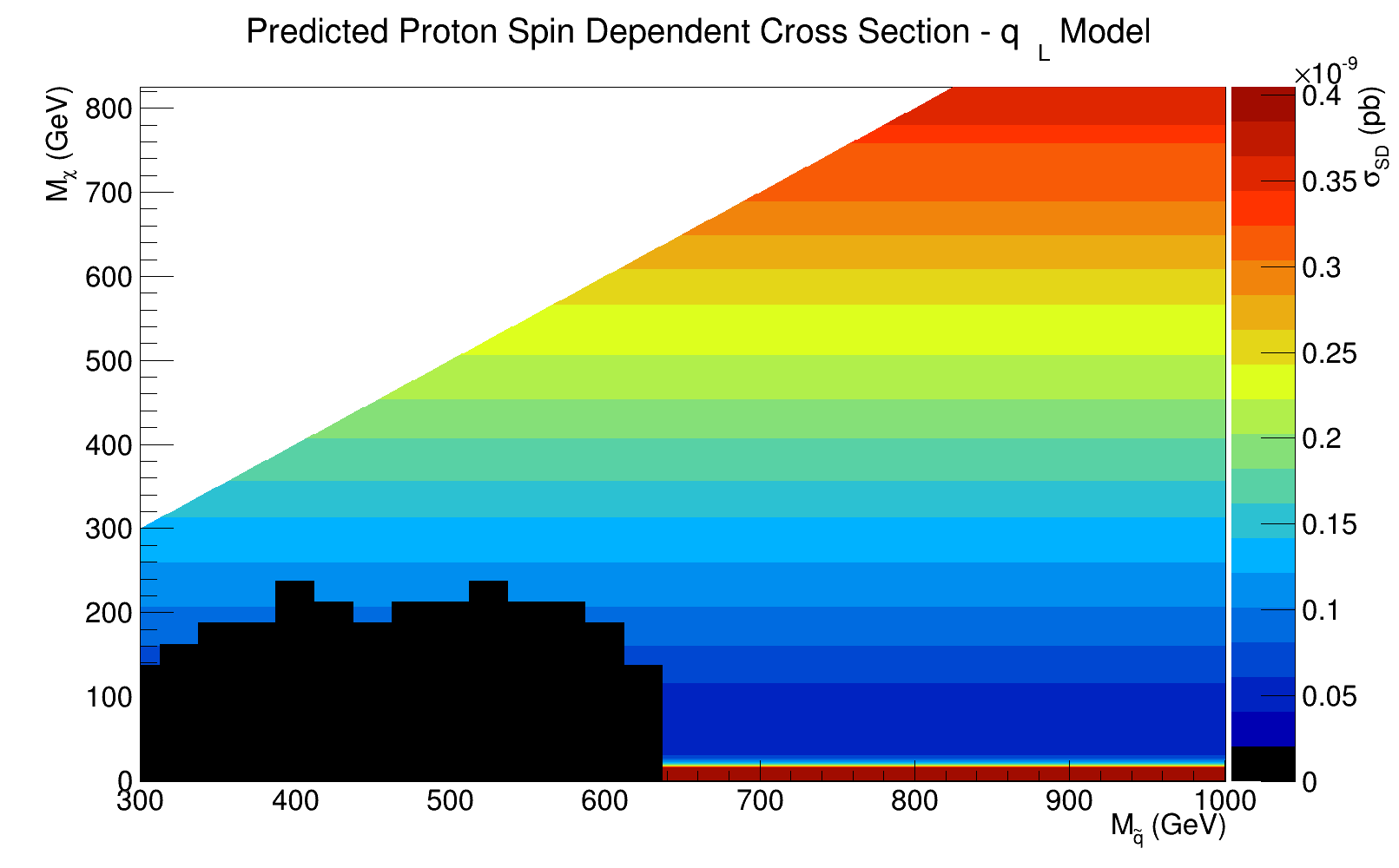}
    \caption{}
    \label{fig:sdpredp-c}
  \end{subfigure}
  \caption{The predicted maximum spin-dependent proton-DM cross section from the combined Collider and Direct Detection bounds for Dirac Dark Matter}\label{fig:sdpredp}
\end{figure}

\begin{figure}
  \begin{subfigure}[b]{0.46\textwidth}
    \includegraphics[width=\textwidth]{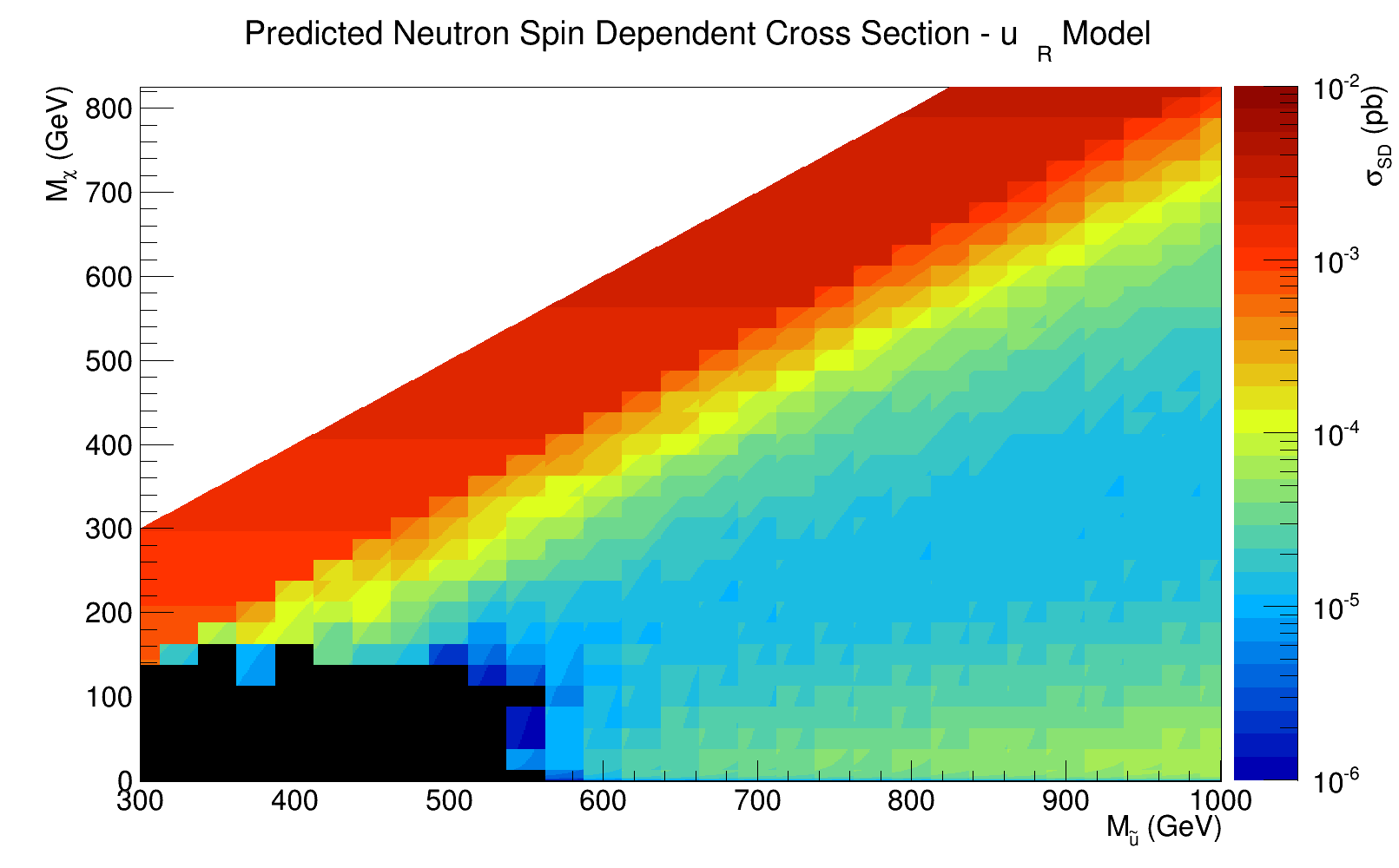}
    \caption{}
    \label{fig:Msdpredn-a}
  \end{subfigure}

  \begin{subfigure}[b]{0.46\textwidth}
    \includegraphics[width=\textwidth]{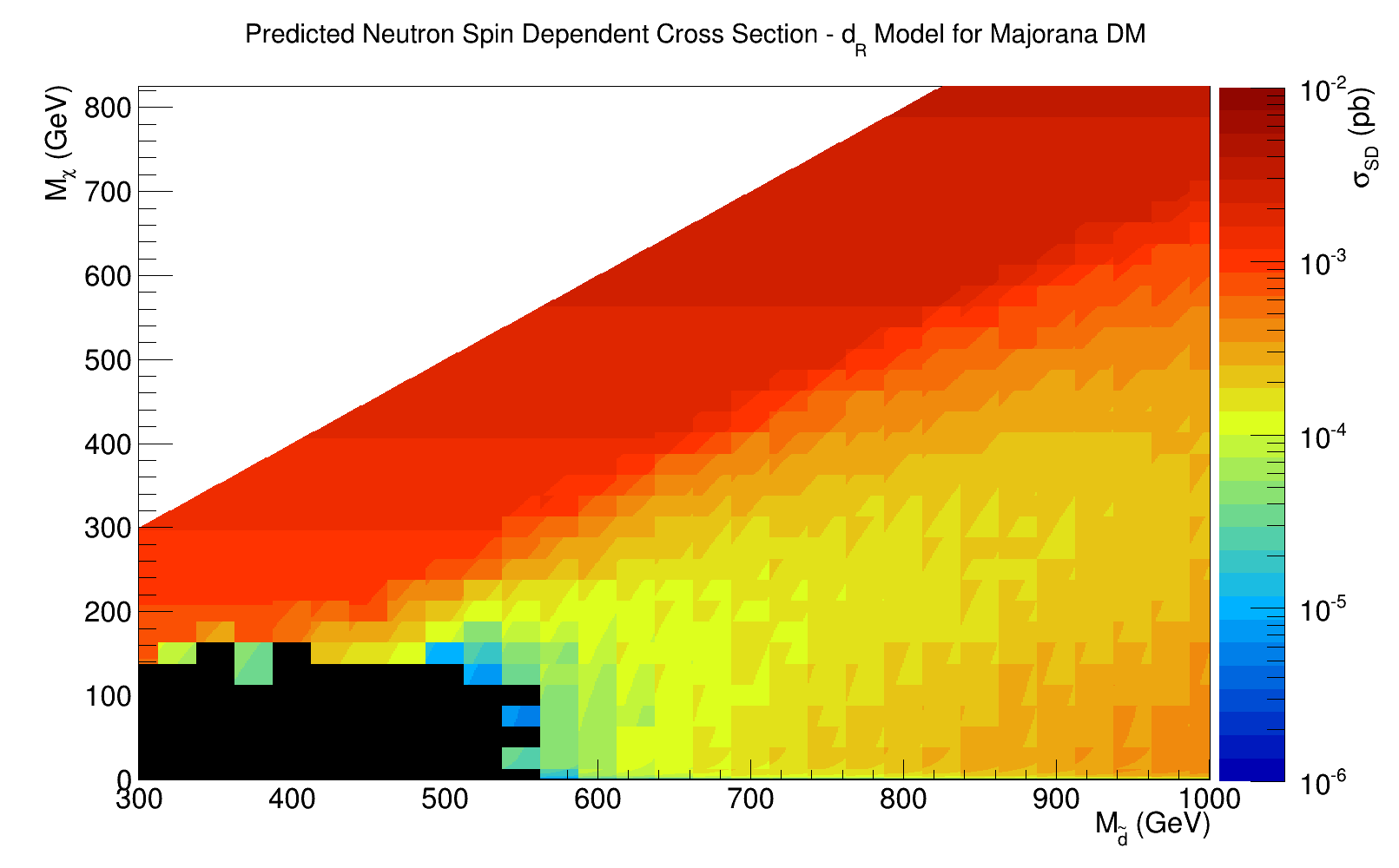}
    \caption{}
    \label{fig:Msdpredn-b}
  \end{subfigure}

  \begin{subfigure}[b]{0.46\textwidth}
    \includegraphics[width=\textwidth]{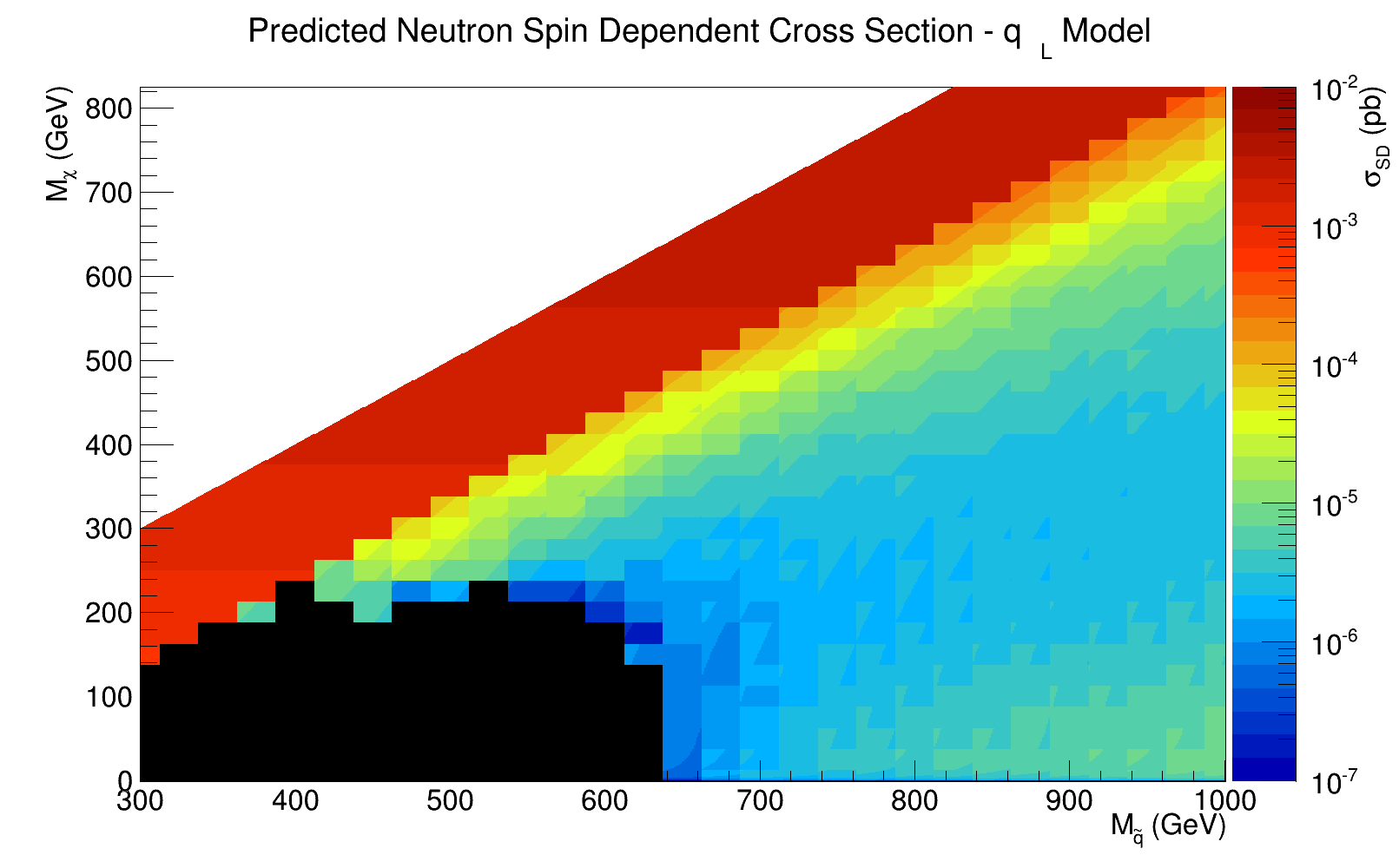}
    \caption{}
    \label{fig:Msdpredn-c}
  \end{subfigure}
  \caption{The predicted maximum spin-dependent neutron-DM cross section from the combined Collider and Direct Detection bounds for Majorana Dark Matter}\label{fig:Msdpredn}
\end{figure}

\begin{figure}
  \begin{subfigure}[b]{0.46\textwidth}
    \includegraphics[width=\textwidth]{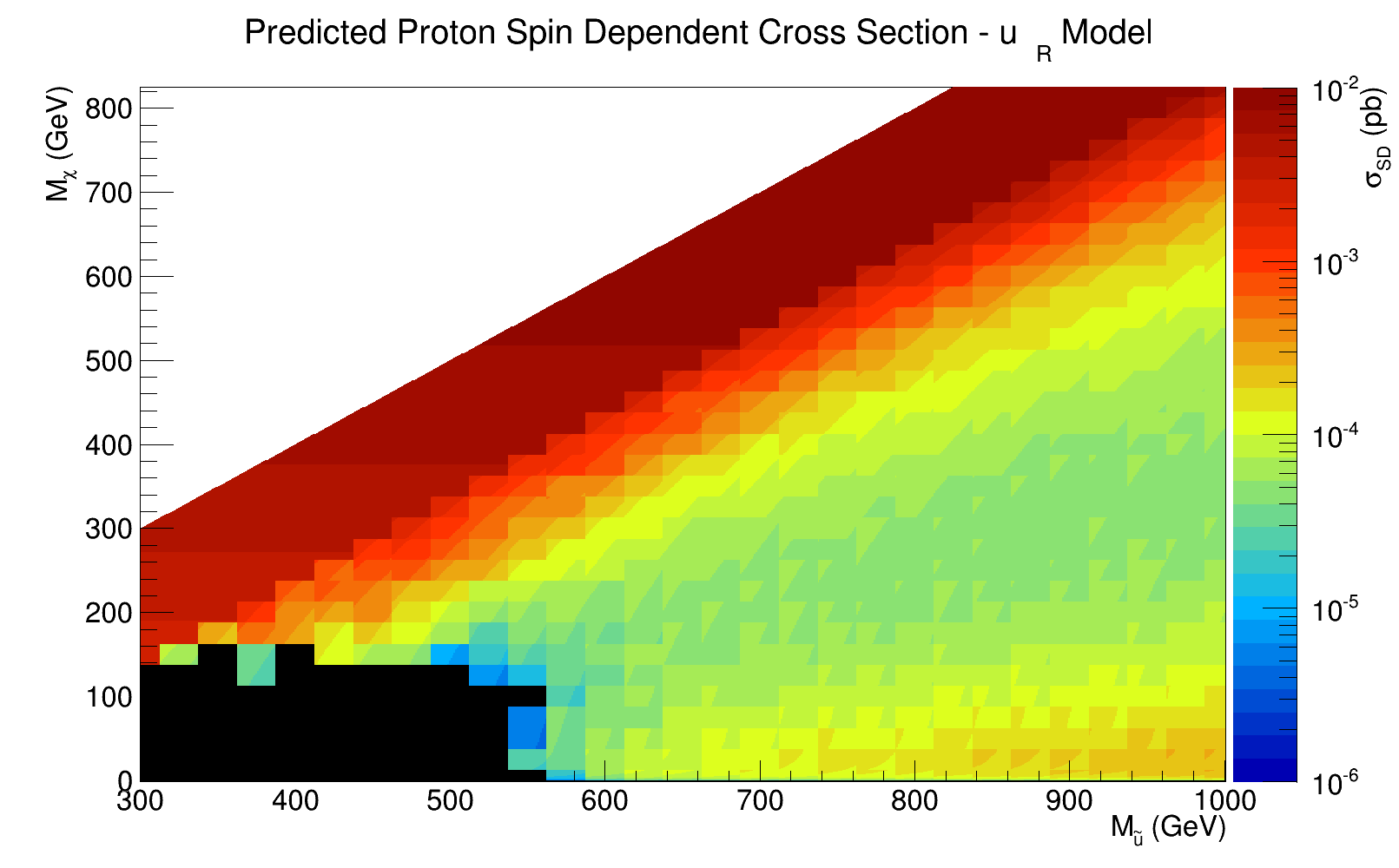}
    \caption{}
    \label{fig:Msdpredp-a}
  \end{subfigure}

  \begin{subfigure}[b]{0.46\textwidth}
    \includegraphics[width=\textwidth]{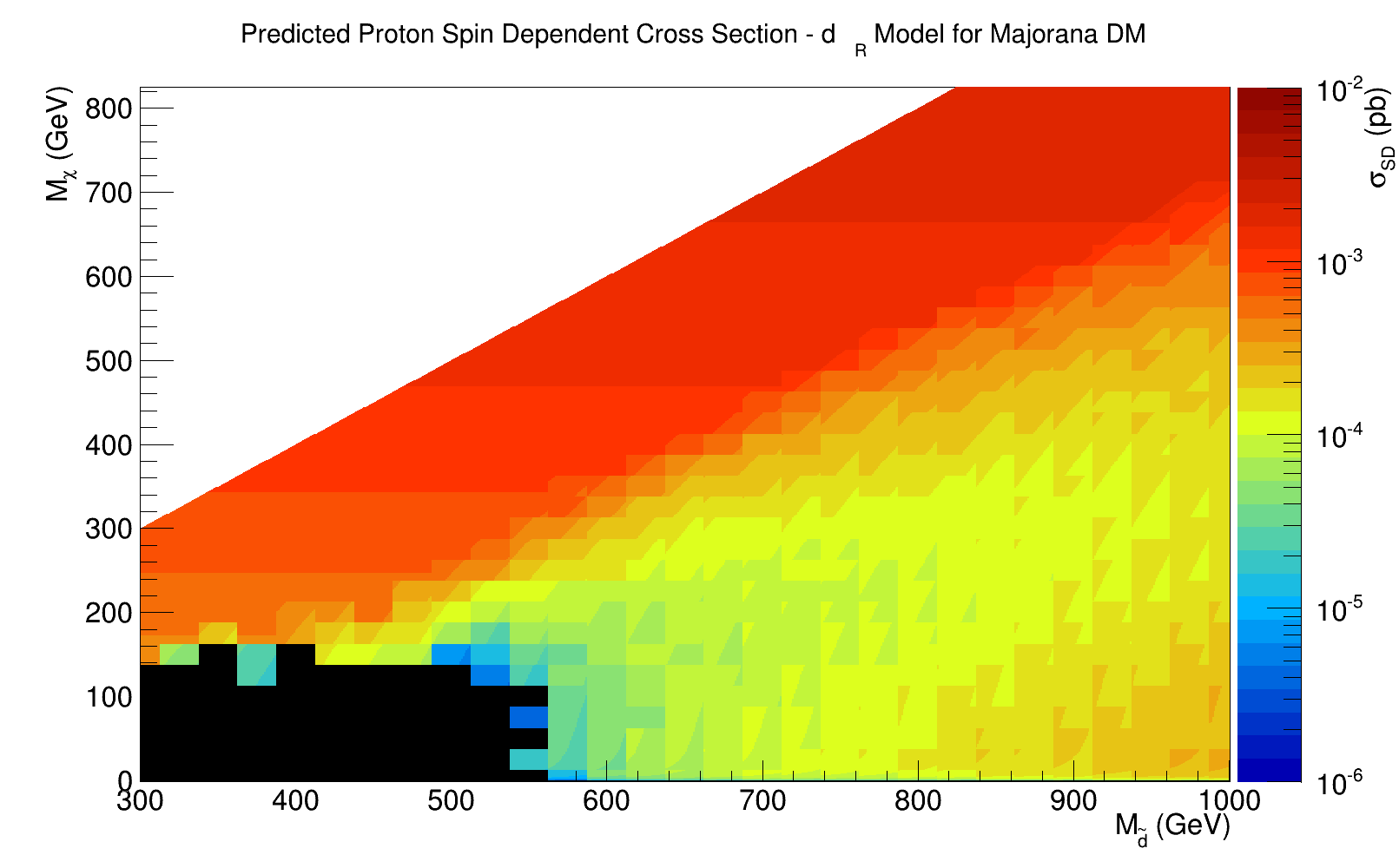}
    \caption{}
    \label{fig:Msdpredp-b}
  \end{subfigure}

  \begin{subfigure}[b]{0.46\textwidth}
    \includegraphics[width=\textwidth]{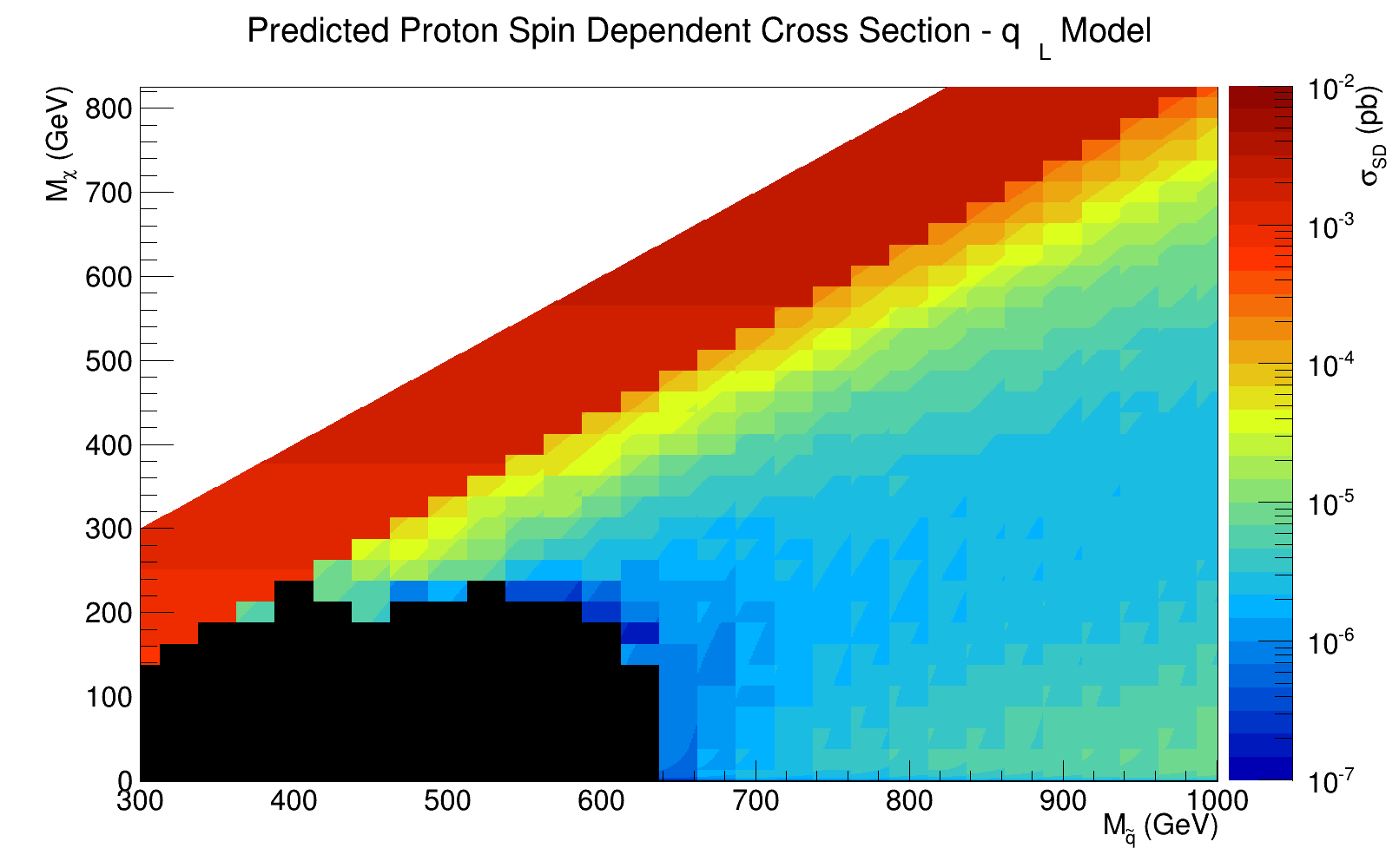}
    \caption{}
    \label{fig:Msdpredp-c}
  \end{subfigure}
  \caption{The predicted maximum spin-dependent proton-DM cross section from the combined Collider and Direct Detection bounds for Majorana Dark Matter}\label{fig:Msdpredp}
\end{figure}

Based on the combined bounds of Figures~\ref{fig:Dbound} and \ref{fig:Mbound}, we can make forecasts
for what other types of experiments might hope to see in this class of simplified model.  For example, for
the Dirac dark matter, there SD interactions which correspond to the maximum values of $g_{\!_{DM}}$
at each mass point.  In Figures~\ref{fig:sdpredn} and \ref{fig:sdpredp}, we show the maximum allowed
spin-dependent cross section on neutrons and protons (respectively), given the constraints from colliders
and XENON.  Based on these results, the outlook for SD searches is such that large improvements over
the current generations of experiments will be necessary if a Dirac simplified model of the type considered here
accurately represents the low energy physics of dark matter.  In contrast, the forecast for SD searches for
Majorana dark matter, shown in Figures~\ref{fig:Msdpredn} and \ref{fig:Msdpredp}, show that there is hope
in the near-future for these searches to make a discovery.

\begin{figure}
  \begin{subfigure}[b]{0.46\textwidth}
    \includegraphics[width=\textwidth]{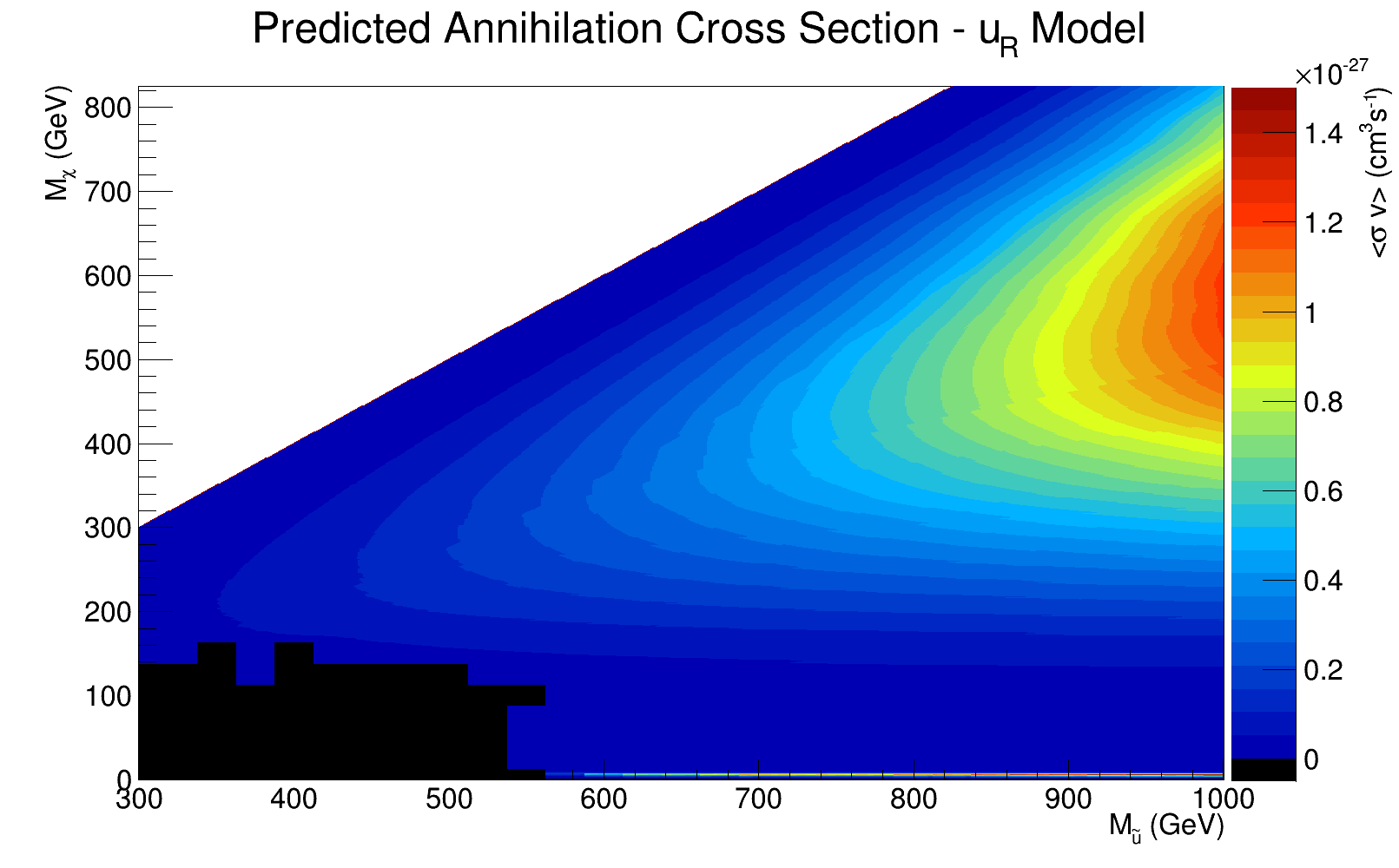}
    \caption{}
    \label{fig:annD-a}
  \end{subfigure}

  \begin{subfigure}[b]{0.46\textwidth}
    \includegraphics[width=\textwidth]{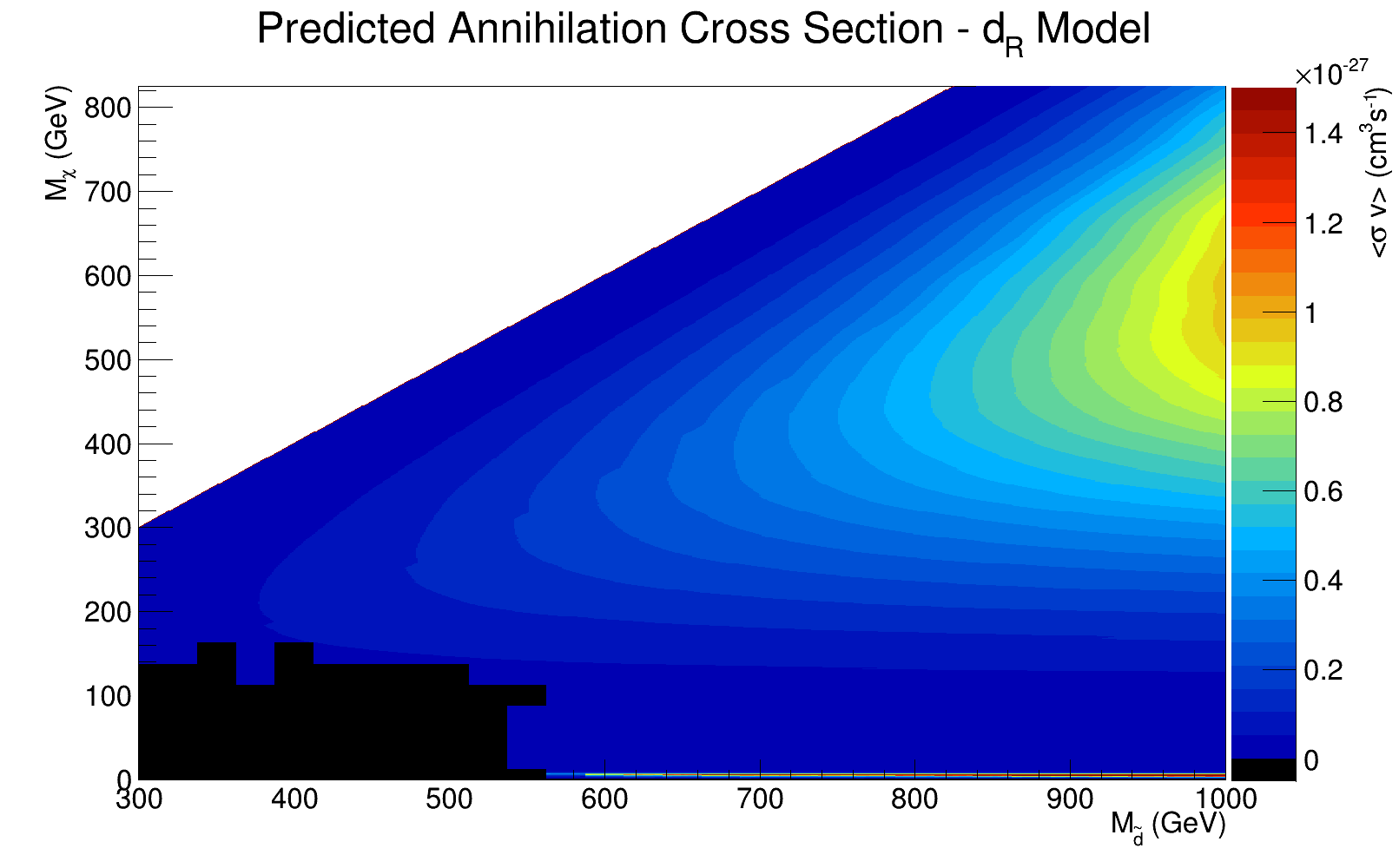}
    \caption{}
    \label{fig:annD-b}
  \end{subfigure}

  \begin{subfigure}[b]{0.46\textwidth}
    \includegraphics[width=\textwidth]{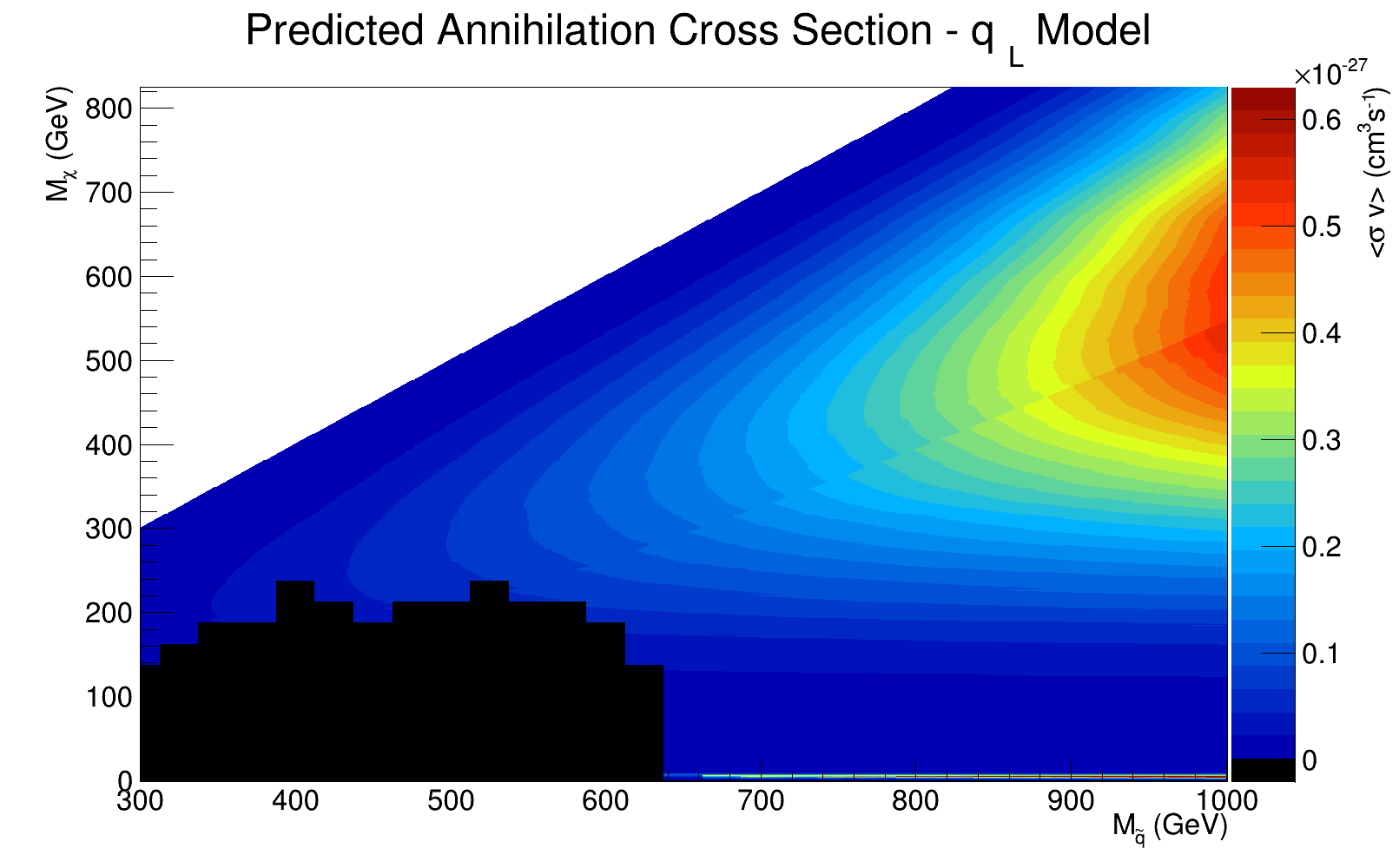}
    \caption{}
    \label{fig:annD-c}
  \end{subfigure}
  \caption{The predicted maximum annihilation cross section from the combined Collider and Direct Detection bounds for Dirac Dark Matter}\label{fig:annD}
\end{figure}

\begin{figure}
  \begin{subfigure}[b]{0.46\textwidth}
    \includegraphics[width=\textwidth]{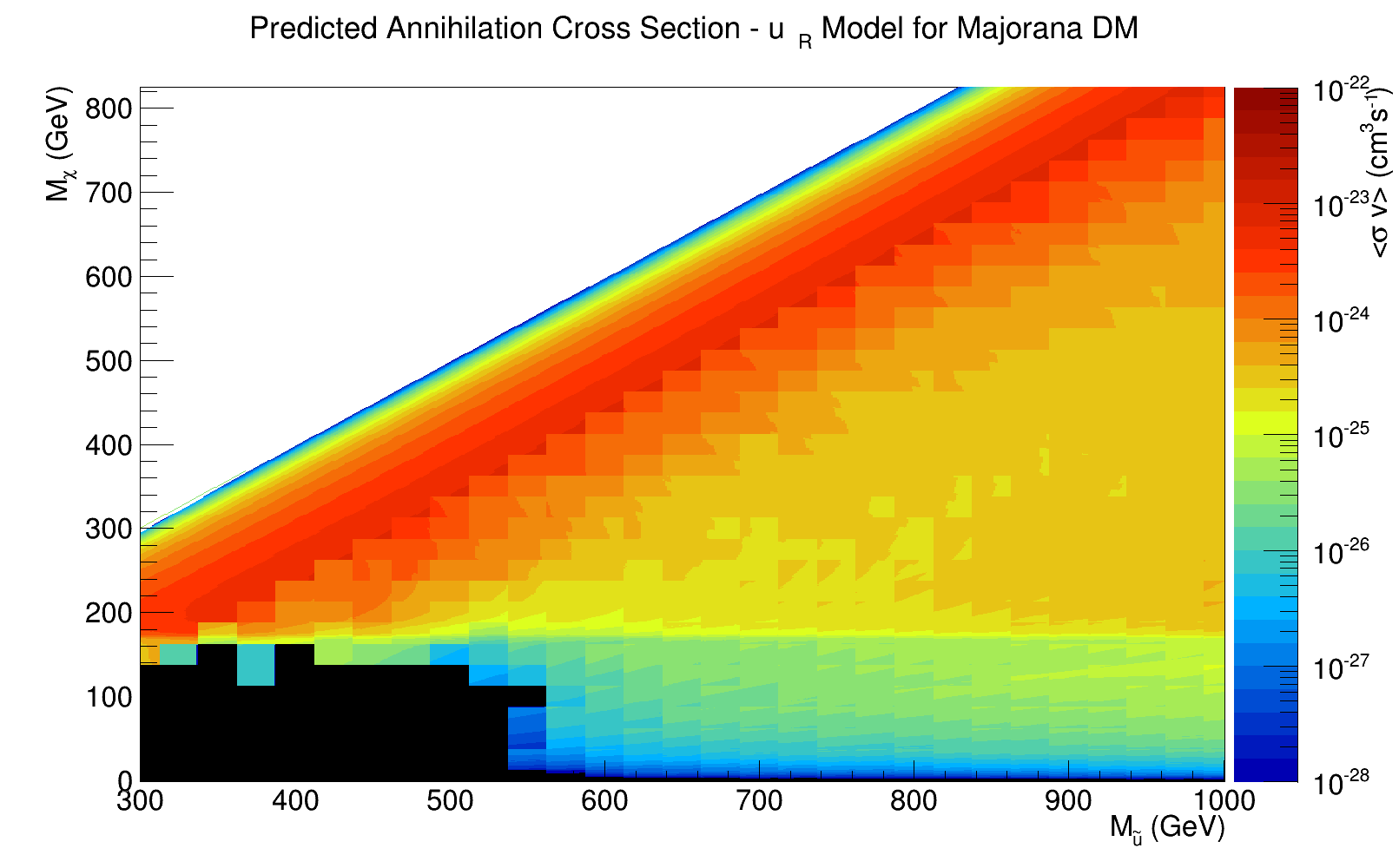}
    \caption{}
    \label{fig:annM-a}
  \end{subfigure}

  \begin{subfigure}[b]{0.46\textwidth}
    \includegraphics[width=\textwidth]{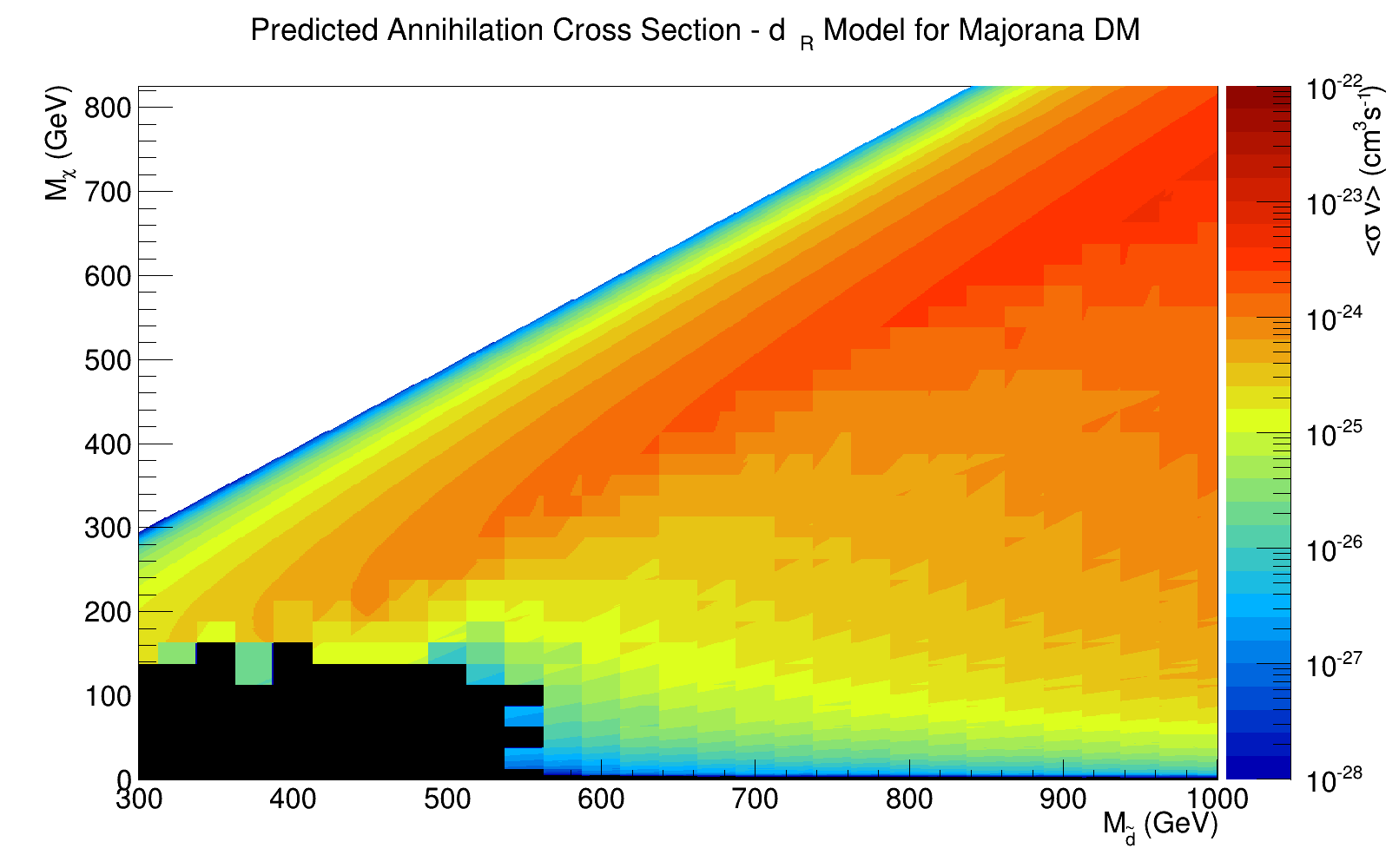}
    \caption{}
    \label{fig:annM-b}
  \end{subfigure}

  \begin{subfigure}[b]{0.46\textwidth}
    \includegraphics[width=\textwidth]{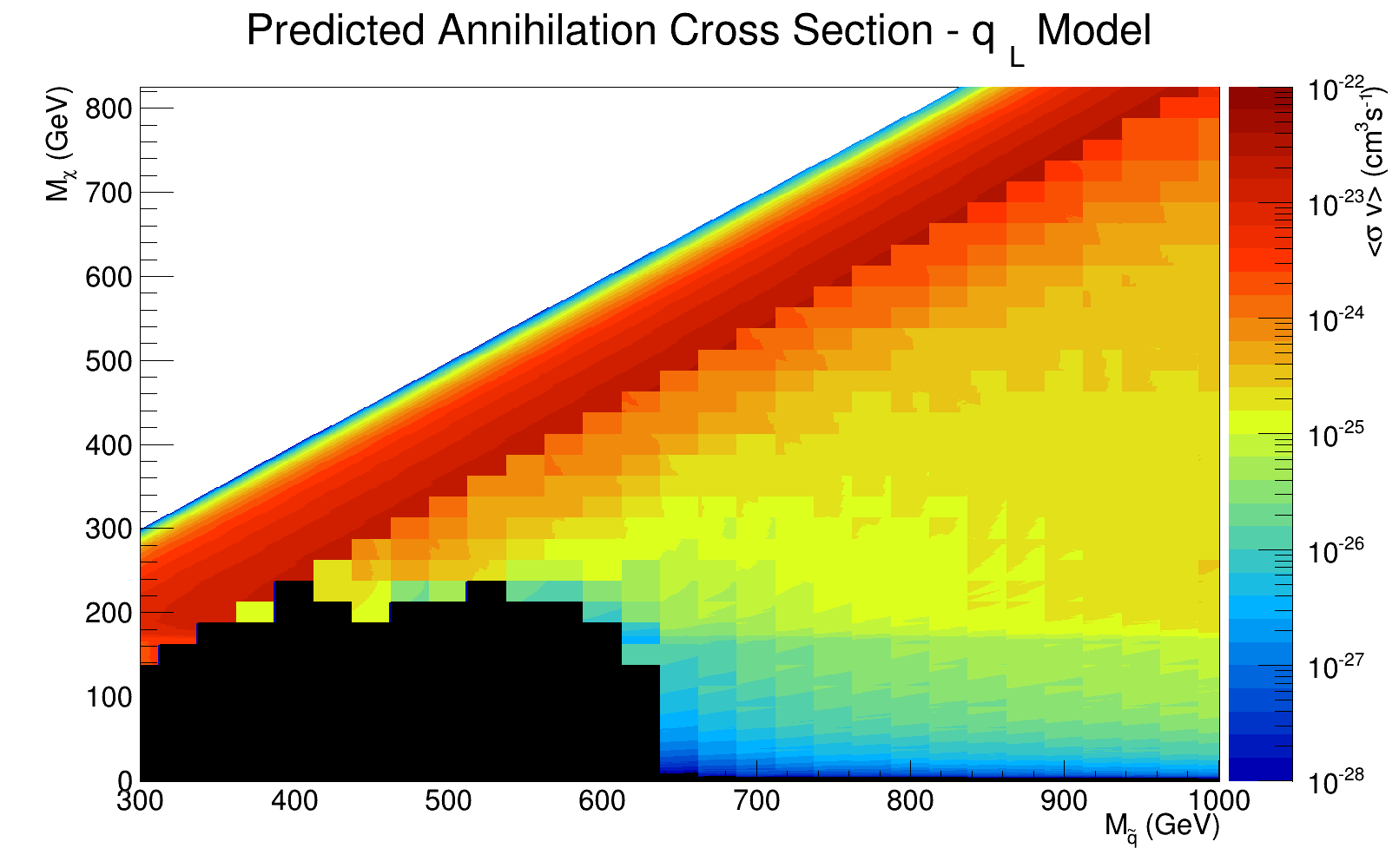}
    \caption{}
    \label{fig:annM-c}
  \end{subfigure}
  \caption{The predicted maximum annihilation cross section from the combined Collider and Direct Detection bounds for Majorana Dark Matter}\label{fig:annM}
\end{figure}

Another very important observable is the annihilation cross section.  
We compute the thermally averaged annihilation cross section to all possible quark pairs 
$\langle \sigma v \rangle$ at
the time of freeze-out using micrOMEGAS \cite{Belanger:2013oya} 
and the results are shown in Figures~\ref{fig:annD}
and \ref{fig:annM}.  These figures indicate that for this class of simplified model, the bounds from colliders
and direct detection are already strong enough that additional annihilation channels would be required
to allow for the dark matter to be a thermal relic with a standard cosmology if it is Dirac.  Instead, if it is
Majorana, there are accessible regions where it could be a standard thermal relic.

\section{Outlook}

We have considered simplified models where the dark matter is either a Dirac or Majorana fermion, and 
interacts with a quark as well as a colored mediator.  
We found that the demands of gauge invariance suggest that when the dark matter is an electroweak singlet,
only three classes of mediator are possible.  Minimal flavor violation further suggests that the couplings
and masses of the mediators are flavor-universal, with some freedom with respect to the third generation
that we chose not to explore here.  It is worth emphasizing that we have been pushed toward a simplified
model that essentially already existed to interpret searches for squark-like states.  We have found that by
trading the LHC production cross section as one of its defining parameters for the strength of
the coupling of the dark matter to a quark and a scalar mediator, we arrive at a full effective field theory
with the same number of parameters, but capable of describing the physics of dark matter even outside
of the context of colliders.

We have derived the bounds from collider searches and elastic scattering of the dark matter from nuclei
via either spin-independent or spin-dependent interactions.  We find that the two probes are complementary
to one another, and provide different kinds of information.   The summarized
bounds for Dirac and Majorana dark matter are shown in Figures~\ref{fig:Dbound} and \ref{fig:Mbound}, which
represent our best knowledge to date of such constructions.  These simplified models are 
potentially UV-complete sketches of dark matter, and can realistically capture some of its most important
features.  If a robust detection of dark matter lies in the near future, the first step to establishing it as dark
matter and putting it into the context of particle physics will be to explore these models and their close
cousins, to determine which ones seem to most accurately explain the experimental data.  Ultimately,
the hope is to connect with a complete theory of dark matter which will extend the SM into a more fundamental
theory.

\section*{Note Added}

In the final stages of preparing this work, 
Refs.~\cite{Chang:2013oia,An:2013xka,Bai:2013iqa} appeared, which have a substantial overlap with
the ideas presented here, though assumptions and presentation have small variations.

\section*{Acknowledgements}

The research of A.R. is
supported in part by the National Science Foundation under grant PHY-0970173
and that of T.M.P.T. is supported in part by NSF
grant PHY-0970171 and by the University of California, Irvine through a Chancellor's fellowship.
The work of K.I.N was supported by Grant-in-Aid for Scientific research from the Ministry of Education, Science, Sports, and Culture (MEXT), Japan (No.23104006), and the JSPS Institutional Program for Young Researcher Overseas Visits of KEK. Part of numerical computation was carried out at the Yukawa Institute Computer Facility. K.I.N would like to thank the particle theory group at UC Irvine for their warm hospitality while this work was initiated.

\bibliographystyle{unsrt}

\begin{thebibliography}{9}

\bibitem{Bauer:2013ihz} 
  D.~Bauer, J.~Buckley, M.~Cahill-Rowley, R.~Cotta, A.~Drlica-Wagner, J.~Feng, S.~Funk and J.~Hewett {\it et al.},
  arXiv:1305.1605 [hep-ph].

\bibitem{Cahill-Rowley:2013dpa} 
  M.~Cahill-Rowley, R.~Cotta, A.~Drlica-Wagner, S.~Funk, J.~Hewett, A.~Ismail, T.~Rizzo and M.~Wood,
  arXiv:1305.6921 [hep-ph].
  
\bibitem{Alves:2011wf} 
  D.~Alves {\it et al.}  [LHC New Physics Working Group Collaboration],
  J.\ Phys.\ G {\bf 39}, 105005 (2012)
  [arXiv:1105.2838 [hep-ph]].

\bibitem{Beltran:2010ww} 
  M.~Beltran, D.~Hooper, E.~W.~Kolb, Z.~A.~C.~Krusberg and T.~M.~P.~Tait,
  JHEP {\bf 1009}, 037 (2010)
  [arXiv:1002.4137 [hep-ph]];
  Q.~-H.~Cao, C.~-R.~Chen, C.~S.~Li and H.~Zhang,
  JHEP {\bf 1108}, 018 (2011)
  [arXiv:0912.4511 [hep-ph]];
  J.~Goodman, M.~Ibe, A.~Rajaraman, W.~Shepherd, T.~M.~P.~Tait and H.~-B.~Yu,
  Phys.\ Lett.\ B {\bf 695}, 185 (2011)
  [arXiv:1005.1286 [hep-ph]];
  Y.~Bai, P.~J.~Fox and R.~Harnik,
  JHEP {\bf 1012}, 048 (2010)
  [arXiv:1005.3797 [hep-ph]];
  J.~Goodman, M.~Ibe, A.~Rajaraman, W.~Shepherd, T.~M.~P.~Tait and H.~-B.~Yu,
  Phys.\ Rev.\ D {\bf 82}, 116010 (2010)
  [arXiv:1008.1783 [hep-ph]];
  P.~J.~Fox, R.~Harnik, J.~Kopp and Y.~Tsai,
  Phys.\ Rev.\ D {\bf 84}, 014028 (2011)
  [arXiv:1103.0240 [hep-ph]];
  A.~Rajaraman, W.~Shepherd, T.~M.~P.~Tait and A.~M.~Wijangco,
  Phys.\ Rev.\ D {\bf 84}, 095013 (2011)
  [arXiv:1108.1196 [hep-ph]];
  P.~J.~Fox, R.~Harnik, J.~Kopp and Y.~Tsai,
  Phys.\ Rev.\ D {\bf 85}, 056011 (2012)
  [arXiv:1109.4398 [hep-ph]];
  P.~J.~Fox, R.~Harnik, R.~Primulando and C.~-T.~Yu,
  Phys.\ Rev.\ D {\bf 86}, 015010 (2012)
  [arXiv:1203.1662 [hep-ph]];
  J.~-F.~Fortin and T.~M.~P.~Tait,
  Phys.\ Rev.\ D {\bf 85}, 063506 (2012)
  [arXiv:1103.3289 [hep-ph]];
  Y.~Bai and T.~M.~P.~Tait,
  Phys.\ Lett.\ B {\bf 723}, 384 (2013)
  [arXiv:1208.4361 [hep-ph]];
  R.~Ding and Y.~Liao,
  JHEP {\bf 1204}, 054 (2012)
  [arXiv:1201.0506 [hep-ph]];
  R.~C.~Cotta, J.~L.~Hewett, M.~P.~Le and T.~G.~Rizzo,
  arXiv:1210.0525 [hep-ph];
  H.~Dreiner, D.~Schmeier and J.~Tattersall,
  Europhys.\ Lett.\  {\bf 102}, 51001 (2013)
  [arXiv:1303.3348 [hep-ph]];
  L.~M.~Carpenter, A.~Nelson, C.~Shimmin, T.~M.~P.~Tait and D.~Whiteson,
  arXiv:1212.3352 [hep-ex];
  Z.~-H.~Yu, Q.~-S.~Yan and P.~-F.~Yin,
  arXiv:1307.5740 [hep-ph].
  
\bibitem{Beltran:2008xg} 
  M.~Beltran, D.~Hooper, E.~W.~Kolb and Z.~C.~Krusberg,
  Phys.\ Rev.\ D {\bf 80}, 043509 (2009)
  [arXiv:0808.3384 [hep-ph]];
  A.~Kurylov and M.~Kamionkowski,
  Phys.\ Rev.\ D {\bf 69}, 063503 (2004)
  [hep-ph/0307185];
  K.~Cheung, P.~-Y.~Tseng, Y.~-L.~S.~Tsai and T.~-C.~Yuan,
  JCAP {\bf 1205}, 001 (2012)
  [arXiv:1201.3402 [hep-ph]];
  J.~Goodman, M.~Ibe, A.~Rajaraman, W.~Shepherd, T.~M.~P.~Tait and H.~-B.~Yu,
  Nucl.\ Phys.\ B {\bf 844}, 55 (2011)
  [arXiv:1009.0008 [hep-ph]];
  K.~Cheung, P.~-Y.~Tseng and T.~-C.~Yuan,
  JCAP {\bf 1101}, 004 (2011)
  [arXiv:1011.2310 [hep-ph]];
  K.~Cheung, P.~-Y.~Tseng and T.~-C.~Yuan,
  JCAP {\bf 1106}, 023 (2011)
  [arXiv:1104.5329 [hep-ph]];
  A.~Rajaraman, T.~M.~P.~Tait and D.~Whiteson,
  JCAP {\bf 1209}, 003 (2012)
  [arXiv:1205.4723 [hep-ph]];
  A.~Rajaraman, T.~M.~P.~Tait and A.~M.~Wijangco,
  Phys.\ Dark Univ.\  {\bf 2}, 17 (2013)
  [arXiv:1211.7061 [hep-ph]];
  A.~De Simone, A.~Monin, A.~Thamm and A.~Urbano,
  JCAP {\bf 1302}, 039 (2013)
  [arXiv:1301.1486 [hep-ph]];
  J.~-M.~Zheng, Z.~-H.~Yu, J.~-W.~Shao, X.~-J.~Bi, Z.~Li and H.~-H.~Zhang,
  Nucl.\ Phys.\ B {\bf 854}, 350 (2012)
  [arXiv:1012.2022 [hep-ph]];
  B.~Bellazzini, M.~Cliche and P.~Tanedo,
  arXiv:1307.1129 [hep-ph];
  I.~M.~Shoemaker,
  arXiv:1305.1936 [hep-ph];
  P.~Gondolo, J.~Hisano and K.~Kadota,
  Phys.\ Rev.\ D {\bf 86}, 083523 (2012)
  [arXiv:1205.1914 [hep-ph]];
  J.~M.~Cornell, S.~Profumo and W.~Shepherd,
  Phys.\  Rev.\  D 88, {\bf 015027} (2013)
  [arXiv:1305.4676 [hep-ph]].
  
\bibitem{Goodman:2011jq} 
  J.~Goodman and W.~Shepherd,
  arXiv:1111.2359 [hep-ph];
  M.~T.~Frandsen, F.~Kahlhoefer, S.~Sarkar and K.~Schmidt-Hoberg,
  JHEP {\bf 1109}, 128 (2011)
  [arXiv:1107.2118 [hep-ph]];
  I.~M.~Shoemaker and L.~Vecchi,
  Phys.\ Rev.\ D {\bf 86}, 015023 (2012)
  [arXiv:1112.5457 [hep-ph]];
  H.~An, X.~Ji and L.~-T.~Wang,
  JHEP {\bf 1207}, 182 (2012)
  [arXiv:1202.2894 [hep-ph]];
  M.~T.~Frandsen, F.~Kahlhoefer, A.~Preston, S.~Sarkar and K.~Schmidt-Hoberg,
  JHEP {\bf 1207}, 123 (2012)
  [arXiv:1204.3839 [hep-ph]];
  G.~Busoni, A.~De Simone, E.~Morgante and A.~Riotto,
  arXiv:1307.2253 [hep-ph];
  R.~C.~Cotta, A.~Rajaraman, T.~M.~P.~Tait and A.~M.~Wijangco,
  arXiv:1305.6609 [hep-ph];
  S.~Profumo, W.~Shepherd and T.~Tait,
  arXiv:1307.6277 [hep-ph].
  Y.~Gershtein, F.~Petriello, S.~Quackenbush and K.~M.~Zurek,
  Phys.\ Rev.\ D {\bf 78}, 095002 (2008)
  [arXiv:0809.2849 [hep-ph]];
  F.~J.~Petriello, S.~Quackenbush and K.~M.~Zurek,
  Phys.\ Rev.\ D {\bf 77}, 115020 (2008)
  [arXiv:0803.4005 [hep-ph]];
  P.~Agrawal, Z.~Chacko, C.~Kilic and R.~K.~Mishra,
  arXiv:1003.1912 [hep-ph].
  M.~Garny, A.~Ibarra, M.~Pato and S.~Vogl,
  JCAP {\bf 1211}, 017 (2012)
  [arXiv:1207.1431 [hep-ph]].
  
\bibitem{D'Ambrosio:2002ex} 
  G.~D'Ambrosio, G.~F.~Giudice, G.~Isidori and A.~Strumia,
  Nucl.\ Phys.\ B {\bf 645}, 155 (2002)
  [hep-ph/0207036].

\bibitem{CMS:zxa} 
  [CMS Collaboration],
  CMS-PAS-SUS-12-028.

\bibitem{Alwall:2011uj} 
  J.~Alwall, M.~Herquet, F.~Maltoni, O.~Mattelaer and T.~Stelzer,
  JHEP {\bf 1106}, 128 (2011)
  [arXiv:1106.0522 [hep-ph]].

\bibitem{Christensen:2008py} 
  N.~D.~Christensen and C.~Duhr,
  Comput.\ Phys.\ Commun.\  {\bf 180}, 1614 (2009)
  [arXiv:0806.4194 [hep-ph]].
  
\bibitem{Beenakker:1996ed} 
  W.~Beenakker, R.~Hopker and M.~Spira,
  hep-ph/9611232.

\bibitem{Nieves:2003in} 
  J.~F.~Nieves and P.~B.~Pal,
  Am.\ J.\ Phys.\  {\bf 72}, 1100 (2004)
  [hep-ph/0306087].

\bibitem{Freytsis:2010ne} 
  M.~Freytsis and Z.~Ligeti,
  Phys.\ Rev.\ D {\bf 83}, 115009 (2011)
  [arXiv:1012.5317 [hep-ph]].
  
\bibitem{Gondolo:2004sc} 
  P.~Gondolo, J.~Edsjo, P.~Ullio, L.~Bergstrom, M.~Schelke and E.~A.~Baltz,
  JCAP {\bf 0407}, 008 (2004)
  [astro-ph/0406204].

\bibitem{Aprile:2012nq} 
  E.~Aprile {\it et al.}  [XENON100 Collaboration],
  Phys.\ Rev.\ Lett.\  {\bf 109}, 181301 (2012)
  [arXiv:1207.5988 [astro-ph.CO]].

\bibitem{Angle:2011th} 
  J.~Angle {\it et al.}  [XENON10 Collaboration],
  Phys.\ Rev.\ Lett.\  {\bf 107}, 051301 (2011)
  [arXiv:1104.3088 [astro-ph.CO]].

\bibitem{Aprile:2013doa} 
  E.~Aprile {\it et al.}  [XENON100 Collaboration],
  arXiv:1301.6620 [astro-ph.CO].

\bibitem{Archambault:2012pm} 
  S.~Archambault {\it et al.}  [PICASSO Collaboration],
  Phys.\ Lett.\ B {\bf 711}, 153 (2012)
  [arXiv:1202.1240 [hep-ex]].
  
\bibitem{Belanger:2013oya} 
  G.~Belanger, F.~Boudjema, A.~Pukhov and A.~Semenov,
  arXiv:1305.0237 [hep-ph].

\bibitem{Chang:2013oia} 
  S.~Chang, R.~Edezhath, J.~Hutchinson and M.~Luty,
  arXiv:1307.8120 [hep-ph].
  
\bibitem{An:2013xka} 
  H.~An, L.~-T.~Wang and H.~Zhang,
  arXiv:1308.0592 [hep-ph].
  
\bibitem{Bai:2013iqa} 
  Y.~Bai and J.~Berger,
  arXiv:1308.0612 [hep-ph].

\end{thebibliography}

\end{document}